\documentclass{aa}
\usepackage[varg]{txfonts}
\usepackage{natbib}
\usepackage{longtable}
\usepackage{pdflscape}
\usepackage{graphicx}
\usepackage{url}
\bibpunct{(}{)}{;}{a}{}{,} 
\usepackage[pdfborderstyle={/S/B}]{hyperref}

\begin{document} 

\title{The \textit{Fermi} GBM gamma-ray burst time-resolved spectral catalog: brightest bursts in the first four years}

\author{Hoi-Fung Yu\inst{\ref{inst1},\ref{inst2}} \and Robert D. Preece\inst{\ref{inst3}} \and Jochen Greiner\inst{\ref{inst1},\ref{inst2}} \and P. Narayana Bhat\inst{\ref{inst4}} \and Elisabetta Bissaldi\inst{\ref{inst5}} \and Michael S. Briggs\inst{\ref{inst4},\ref{inst6}} \and William H. Cleveland\inst{\ref{inst7}}  \and Valerie Connaughton\inst{\ref{inst4},\ref{inst6}} \and Adam Goldstein\inst{\ref{inst8}} \and Andreas von Kienlin\inst{\ref{inst1}} \and Chryssa Kouveliotou\inst{\ref{inst9}} \and Bagrat Mailyan\inst{\ref{inst4}} \and Charles A. Meegan\inst{\ref{inst7}} \and William S. Paciesas\inst{\ref{inst7}} \and Arne Rau\inst{\ref{inst1}} \and Oliver J. Roberts\inst{\ref{inst10}} \and P{\'e}ter Veres\inst{\ref{inst4}} \and Colleen Wilson-Hodge\inst{\ref{inst8}} \and Bin-Bin Zhang\inst{\ref{inst4}} \and Hendrik J. van Eerten\inst{\ref{inst1}}\thanks{Fellow of the Alexander v. Humboldt Foundation}}

\institute{Max-Planck-Institut f{\"u}r extraterrestrische Physik, Giessenbachstra{\ss}e 1, 85748 Garching, Germany\\ \email{sptfung@mpe.mpg.de}\label{inst1}
\and Excellence Cluster Universe, Technische Universit{\"a}t M{\"u}nchen, Boltzmannstra{\ss}e 2, 85748 Garching, Germany\label{inst2}
\and Space Science Department, University of Alabama in Huntsville, Huntsville, AL 35809, USA\label{inst3}
\and Center for Space Plasma and Aeronomic Research, University of Alabama in Huntsville, Huntsville, AL 35805, USA\label{inst4}
\and INFN, Sez. di Bari, Via E. Orabona 4, 70125 Bari, Italy\label{inst5}
\and Physics Department, University of Alabama in Huntsville, Huntsville, AL 35805, USA\label{inst6}
\and Universities Space Research Association, Huntsville, AL 35805, USA\label{inst7}
\and Astrophysics Office, ZP12, NASA/Marshall Space Flight Center, Huntsville, AL 35812, USA\label{inst8}
\and Physics Department, George Washington University, 725 21st Street NW, Washington, DC 20052, USA\label{inst9}
\and School of Physics, University College Dublin, Belfield, Dublin 4, Ireland\label{inst10}
}

\abstract
{}
{We aim to obtain high-quality time-resolved spectral fits of gamma-ray bursts (GRBs) observed by the Gamma-ray Burst Monitor (GBM) on board the \textit{Fermi} Gamma-ray Space Telescope.}
{We perform time-resolved spectral analysis with high temporal and spectral resolution of the brightest bursts observed by \textit{Fermi} GBM in its first 4 years of mission.}
{We present the complete catalog containing 1,491 spectra from 81 bursts with high spectral and temporal resolution. Distributions of parameters, statistics of the parameter populations, parameter-parameter and parameter-uncertainty correlations, and their exact values are obtained and presented as main results in this catalog. We report a criterion that is robust enough to automatically distinguish between different spectral evolutionary trends between bursts. We also search for plausible blackbody emission components and find that only 3 bursts (36 spectra in total) show evidence of a pure Planck function. It is observed that the averaged time-resolved low-energy power-law index and peak energy are slightly harder than the time-integrated values. Time-resolved spectroscopic results should be used when interpreting physics from the observed spectra, instead of the time-integrated results.}
{}
{}


\titlerunning{\textit{Fermi} GBM GRB time-resolved spectral catalog}
\maketitle

\section{Introduction}

Gamma-ray bursts (GRBs) are the most energetic explosions known to humankind. Although discovered in 1967 \citep{Klebesadel73a} by the \textit{Vela} Satellite Network, the physics of GRBs still remains unsolved. For example, the exact nature of the emission mechanism of the so-called prompt emission phase is still unclear. Today, we know that GRBs are gamma-ray emissions from cosmological sources \citep{Metzger97a} distributed isotropically across the sky \citep{Meegan92a,Pendleton94a,Briggs96a}. The two kinds of GRBs, long/soft and short/hard \citep{Kouveliotou93a}, are thought to have different origins. It is generally believed that long/soft (duration $T_{90} > 2$~s and low-energy photon rich) GRBs are the result of gravitational collapse events from massive progenitors, and short/hard ($T_{90} < 2$~s and high-energy photon rich) GRBs originate from compact merger events.

A powerful method to discern the physical properties and emission mechanisms of GRBs is through detailed spectral analysis. However, the spectral properties of individual GRB may be significantly different. Therefore, analysis of large samples of burst spectra is necessary to obtain a coherent physical picture. Such large spectral catalogs, some time-integrated and some time-resolved, have been published for many hard X-ray or gamma-ray observing instruments, e.g., the \textit{CGRO}/BATSE \citep[25~keV - 2~MeV,][]{Pendleton94a,Preece00a,Kaneko06a,Goldstein13a}, \textit{BeppoSAX}/GRBM \citep[40 - 700~keV,][]{Frontera09a}, \textit{Swift}/XRT \citep[0.2 - 10~keV,][]{Evans09a}, \textit{Swift}/BAT \citep[15 - 150~keV,][]{Sakamoto08a,Sakamoto11a}, \textit{Fermi}/LAT \citep[20~MeV - 300~GeV,][]{Ackermann13a}, and \textit{Fermi}/GBM \citep[time-integrated, 8~keV - 40~MeV,][]{Nava11a,Goldstein12a,Gruber14a,vonKienlin14a}.

This paper presents the first \textit{Fermi} Gamma-ray Burst Monitor (GBM) gamma-ray burst time-resolved spectral catalog. In contrast to previous time-resolved catalogs of other instruments, the broad energy range covered by the GBM allows a sensitive investigation at energies of a few hundred keV where the peaks or breaks of the prompt emission spectra are located. This catalog presents time-resolved fit parameters using standard fit functions, parameter-parameter and parameter-uncertainty correlations, spectral evolutionary trends over time (in particular the peak energy $E_\text{p}$ evolution), distributions of spectral slopes (given in photon indices $\alpha$ and $\beta$), and plausible blackbody components. A novel measure of the sharpness of the spectral peak has been reported separately for the same burst sample by \citet{Yu15b}. The measure places a strong constraint on the physics of prompt emission models, ruling out an optically thin synchrotron origin for the peak or break of the spectrum in a large majority of cases.

This paper is structured as follows. We describe the characteristics of GBM and the methods of data selection and reduction in Section~\ref{sect:method}. The fitting models used in this catalog are described in Section~\ref{sect:models}. The spectral analysis procedure is given in Section~\ref{sect:ana}, and the fitting results are presented in Section~\ref{sect:results}. We summarize our results and conclude in Section~\ref{sect:conc}. The spectral fitting results are tabulated in Appendix~\ref{app:bigtable}. Unless otherwise stated, all errors reported in this paper are given at the $1\sigma$ confidence level.

\section{The data}
\label{sect:method}

\subsection{Instrumentation}

The \textit{Fermi} Gamma-ray Space Telescope, launched in June 2008, harbors two scientific instruments, namely the Gamma-ray Burst Monitor \citep[GBM,][]{Meegan09a} and the Large Area Telescope \citep[LAT,][]{Atwood09a}. The GBM covers the energy range from  8~keV to 40~MeV, while the LAT is sensitive in the complementary energy range from 30~MeV to 300~GeV. The GBM observes the whole sky which is not occulted by the Earth ($>8$~sr) and provides real-time locations for GRB triggers. These real-time locations are circulated via the Gamma-ray Coordination Network\footnote{\url{http://gcn.gsfc.nasa.gov/gcn3_archive.html}} (GCN) which allows ground-based follow-up observations. Occasionally, an Autonomous Repoint Request (ARR) can be accepted by the Flight Software (FSW) which allows \textit{Fermi} to slew towards the direction of the source, so that it can be observed with the LAT.

There are twelve thallium activated sodium iodide detectors (NaI(Tl), hereafter NaI) and two bismuth germanate detectors (BGO) in the GBM instrument. These detectors serve as a sensitive scintillation array covering both the softer photons by the NaIs (8 - 900~keV) and the harder photons by the BGOs (250~keV - 40~MeV). The arrangement of the NaI detectors allows GBM to locate GRBs in a real-time manner; and the two BGO detectors are placed on opposite sides of the spacecraft in order to cover all bursts coming from any direction in the sky. The wide spectral coverage of over 3 orders of magnitude is the key to detailed spectral analysis for the GRB prompt emission phase.

\subsection{Detector selection}

We apply the same detector selection criteria used in all official GBM GRB time-integrated spectral catalogs \citep{Goldstein12a,Gruber14a,vonKienlin14a}. The detectors with viewing angle larger than 60\degr or blocked by the LAT or solar panels are removed \citep{Bissaldi09a,Goldstein12a,Gruber14a}. For every spectrum, a maximum of three NaIs with one BGO are used in the analysis. If more than three NaIs satisfied these criteria, the NaIs with the smallest viewing angles are used in order to avoid binning bias towards the lower energies \citep{Goldstein12a}.

\subsection{Data type selection}

There are three different types of data generated by GBM. The first type is CTIME which provides coarse spectral resolution of 8 energy channels and fine temporal resolution of 0.256~s during the nominal time period, i.e., before the trigger and 600~s after the trigger; during the trigger period, the resolution is increased to 64~ms. The second type is CSPEC which provides coarse temporal resolution at nominal (4.096~s) and trigger (1.024~s) period, and high spectral resolution of 128 pseudo-logarithmically scaled energy channels. The third type is time-tagged event (TTE) data which stores individual photon events tagged with arrival time (resolution of 2~$\mu s$), photon energy channel (128 pseudo-logarithmic energy channels), and detector number (NaI 0 - 11 and BGO 0 - 1). TTE data were stored on board GBM in a recycling buffer. After 26 November 2012\footnote{\url{http://fermi.gsfc.nasa.gov/ssc/data/access/gbm/}} this data type became continuous. When GBM is triggered, the spacecraft will transmit pre- and post-trigger TTE data (about 300~s in duration) to the ground as science data.

Since only TTE data from $\sim$30~s pre-trigger until $\sim$300~s post-trigger are available, for the bursts with evident precursor or emission longer than 300~s, CSPEC data (about 8,000~s in duration) are used. In this paper, CSPEC data are used for 15 GRBs, and for all other bursts TTE data are used.

\subsection{Energy channel selection and background fitting}

To account for the poor transparency for gamma rays of the silicone pad in front of the NaI crystal and of the Multi Layer Insulation (MLI) around the detectors \citep{Bissaldi09a}, the energy channels below 8~keV and the overflow channels above 900~keV are removed. A similar cutoff criterion is also used in the BGOs so that only energy channels between 250~keV and 40~MeV are used. An effective energy range from 8~keV to 40~MeV is used for the spectral analysis in this paper.

For each burst, a polynomial with order 2 - 4 is fit to every energy channel according to two user-defined background intervals, before and after the emission period. The background model is then interpolated across the emission period. This is done by varying the selected intervals and order of polynomial until the $\chi^2$ statistics is minimised over all energy channels. The resulting background intervals are then loaded to all detectors, generating the background model to be used in the spectral analysis. The background intervals used in this catalog are identical to those used in \citet{Gruber14a}.

\subsection{Burst and spectrum selection}

We first select all the bursts detected by GBM in the first 4 years (i.e., from 14 July 2008 to 13 July 2012), which is the same GRB subset as used in the 4-yr GBM GRB time-integrated spectral catalog \citep{Gruber14a,vonKienlin14a}. The GBM triggered on 954 GRBs in this period of time \citep[one of them triggered GBM twice, see][]{vonKienlin14a}. Time-resolved spectral analysis requires bright bursts with sufficiently high signal-to-noise spectra. This bright subsample is selected by applying the following criteria: 10~keV - 1~MeV energy fluence $f>4 \times 10^{-5}$~erg~cm$^{-2}$ and/or 10~keV - 1~MeV peak photon flux $F_\text{p}>20$~ph~s$^{-1}$~cm$^{-2}$ (in either 64, 256, or 1,024~ms binning timescales). These criteria are satisfied by 134 bursts out of the 954. Sixteen among them are of the short burst class.

In order to alleviate the problem that the spectra from the brightest bursts dominate the statistics, we further require each event to have at least 5 time bins in the light curves when binned with signal-to-noise ratio (S/N) $=30$. This optimal S/N is found by iterating the binning process on characteristic bursts drawn from various fluence and peak-flux level, which does not significantly merge peaks and valleys in the light curves while providing the highest number of time bins. As a result, 81 bursts satisfy these criteria; among them there is only one short burst (GRB 120323A; GBM trigger \#120323507). In total, 1,802 time-resolved time bins and spectra were obtained.

Four different empirical models are fit to each spectrum, resulting in a compilation of $1,802 \times 4 = 7,208$ spectral fits. Compared to the 4-yr GBM GRB time-integrated spectral catalog \citep{Gruber14a,vonKienlin14a}, the catalog presented here includes a lower number of GRBs (81 vs. 943), however, the number of high-resolution spectra is higher (1,491 BEST model fits, see Sect.~\ref{sect:results}, vs. 943).

\section{Fitting models}
\label{sect:models}

Four different empirical models are fit to the spectra in our sample, namely, the Band function, a smoothly broken power law, a cutoff power law (aka. the Comptonized model), and a simple power law.

\subsection{The Band function}

The Band function (BAND) is a model in which a power law with high-energy exponential cutoff and a high-energy power law are joined together by a smooth transition. It is an empirical function proposed by \citet{Band93a}, which fits most of the observed GRB spectra. Parametrized by the peak energy $E_\text{p}$ (despite the fact that there may not be a peak in the $\nu F_\nu$ space when the high-energy photon index $\beta \geq -2$) in the observed $\nu F_\nu$ spectrum, the photon model of BAND is defined as
\begin{equation}
\label{eqn:band}
f_\text{BAND}(E) = A\left\{
\begin{array}{ll}
	\left(\frac{E}{100\text{ keV}}\right)^\alpha \exp\left[-\frac{(\alpha+2)E}{E_\text{p}}\right]: E<E_\text{c} \, , \\
	\left(\frac{E}{100\text{ keV}}\right)^\beta \exp\left(\beta-\alpha\right) \left(\frac{E_\text{c}}{100\text{ keV}}\right)^{\alpha-\beta}: E\geq E_\text{c} \, ,
\end{array}
\right.
\end{equation}
where
\begin{equation}
\label{eqn:Ec}
E_\text{c}=\left(\frac{\alpha-\beta}{\alpha+2}\right)E_\text{p} \, .
\end{equation}
In Eqns.~(\ref{eqn:band}) and (\ref{eqn:Ec}), $A$ is the normalization factor at 100~keV in units of ph~s$^{-1}$~cm$^{-2}$~keV$^{-1}$, $\alpha$ is the low-energy power-law photon index, $\beta$ is the high-energy power-law photon index, $E_\text{p}$ is the peak energy in the $\nu F_\nu$ space in units of keV, and $E_\text{c}$ is the characteristic energy in units of keV.

We note that the \textit{peak energy} $E_\text{p}$ represents the position of the peak in the model curve in the $\nu F_\nu$ space, and the \textit{characteristic energy} $E_\text{c}$ represents the position where the low-energy power law with an exponential cutoff ends and the pure high-energy power law starts. These two energies should be distinguished from the \textit{break energy} $E_\text{b}$ which represents the position where the low-energy power law joins the high-energy power law. Therefore, we should not compare the Band function's $E_\text{p}$ or $E_\text{c}$ to the smoothly broken power law's $E_\text{b}$. In order to facilitate a fair comparison of the parameters, we compute the break energy where the two power laws join together for the Band function. The derivation is already given by \citet{Kaneko06a}, here we only give the resulting equation:
\begin{equation}
\label{eqn:bandeb}
E_\text{b}=\left(\frac{\alpha-\beta}{\alpha+2}\right) \frac{E_\text{p}}{2} + 4 \, ,
\end{equation}
in units of keV. We note that the last constant term corresponds to 1/2 of the lower boundary of the detectors, 8~keV for the NaIs in our case. In the asymptotic limit, this term vanishes and therefore $E_\text{b}$ is proportional to $E_\text{p}$.

\subsection{The smoothly broken power law}

The smoothly broken power law (SBPL) is a model of two power laws joined by a smooth transition. It was first parameterized by \citet{Ryde99a} and then re-parameterized by \citet{Kaneko06a}:
\begin{equation}
\label{eqn:sbpl}
f_\text{SBPL}(E) = A \left(\frac{E}{100\text{ keV}}\right)^b 10^{(a-a_\text{piv})} \, ,
\end{equation}
where
\begin{equation}
\label{eqn:sbplpara}
\left\{
\begin{array}{ll}
a = m \Delta \ln\left(\frac{e^q+e^{-q}}{2}\right), a_\text{piv} = m \Delta \ln\left(\frac{e^{q_\text{piv}}+e^{-q_\text{piv}}}{2}\right) \, , \\
m = \frac{\beta-\alpha}{2} \, , b = \frac{\alpha+\beta}{2} \, , \\
q = \frac{\log(E/E_\text{b})}{2} \, , q_\text{piv} = \frac{\log(100\text{ keV}/E_\text{b})}{2} \, .
\end{array}
\right.
\end{equation}
In Eqns.~(\ref{eqn:sbpl}) and (\ref{eqn:sbplpara}), $A$ is the normalization factor at 100~keV in units of ph~s$^{-1}$~cm$^{-2}$~keV$^{-1}$, $\alpha$ and $\beta$ are the low- and high-energy power-law photon indices respectively, $E_\text{b}$ is the break energy in units of keV, and $\Delta$ is the break scale. Unlike the Band function, the break scale is not coupled to the power-law indices, so SBPL is a 5-parameters model if we let $\Delta$ free to vary. We follow \citet{Kaneko06a}, \citet{Goldstein12a}, and \citet{Gruber14a} to fix $\Delta=0.3$.

The peak energy of SBPL in the $\nu F_\nu$ space can be found at
\begin{equation}
\label{eqn:sbplep}
E_\text{p} = 10^x E_\text{b} \, , x = \Delta \tanh^{-1} \left(\frac{\alpha+\beta+4}{\alpha-\beta}\right) \, .
\end{equation}
We note that Eqn.~(\ref{eqn:sbplep}) is only valid for $\alpha>-2$ and $\beta<-2$.

\subsection{The cutoff power law}

\begin{figure}
\resizebox{\hsize}{!}
{\includegraphics{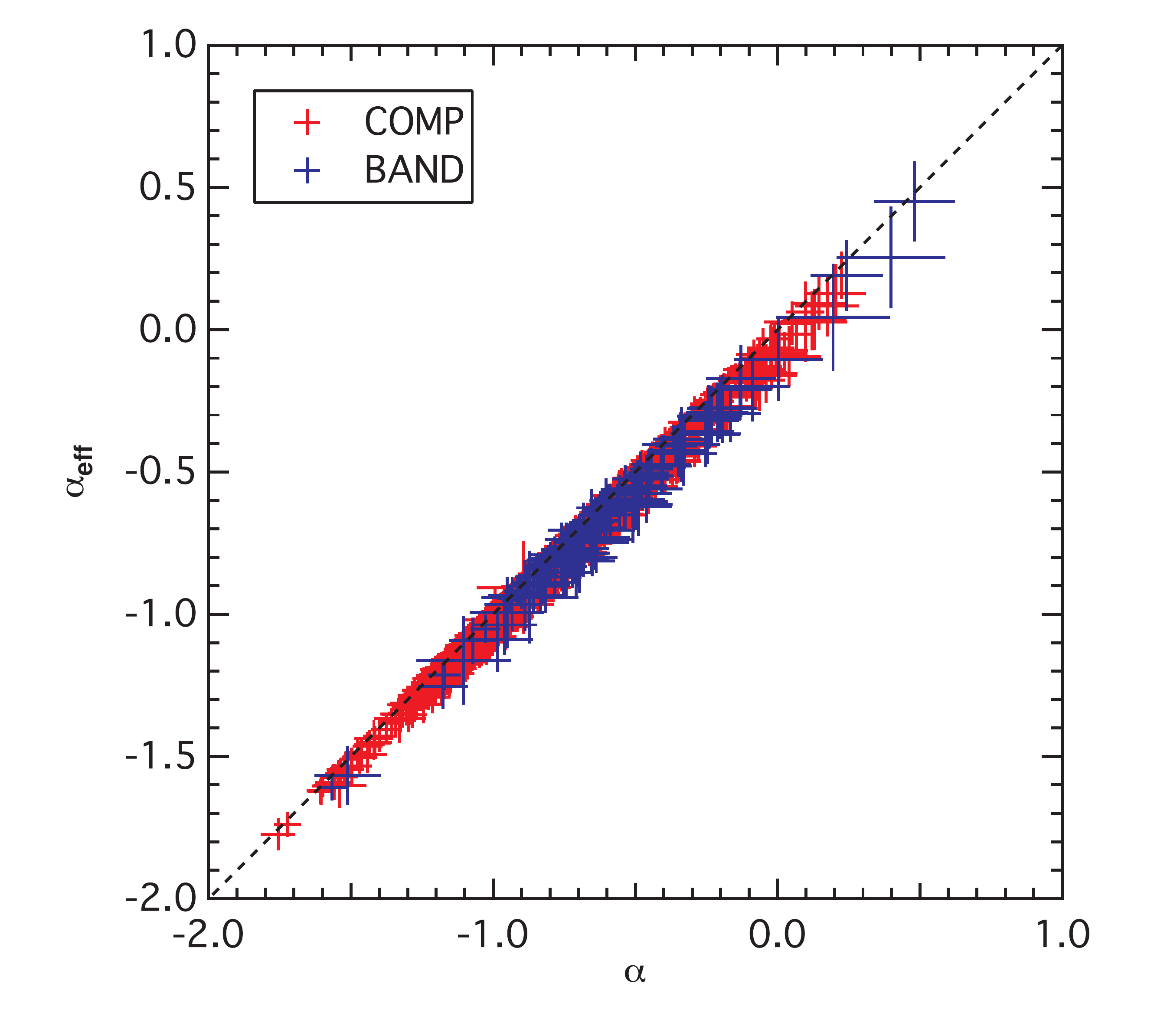}}
\caption{Comparison between $\alpha_\text{eff}$ evaluated at 8~keV and $\alpha$. Blue and red crosses represent BAND and COMP, respectively. Diagonal dashed line shows $y = x$.}
\label{fig:effective}
\end{figure}

\begin{figure}
\resizebox{\hsize}{!}
{\includegraphics{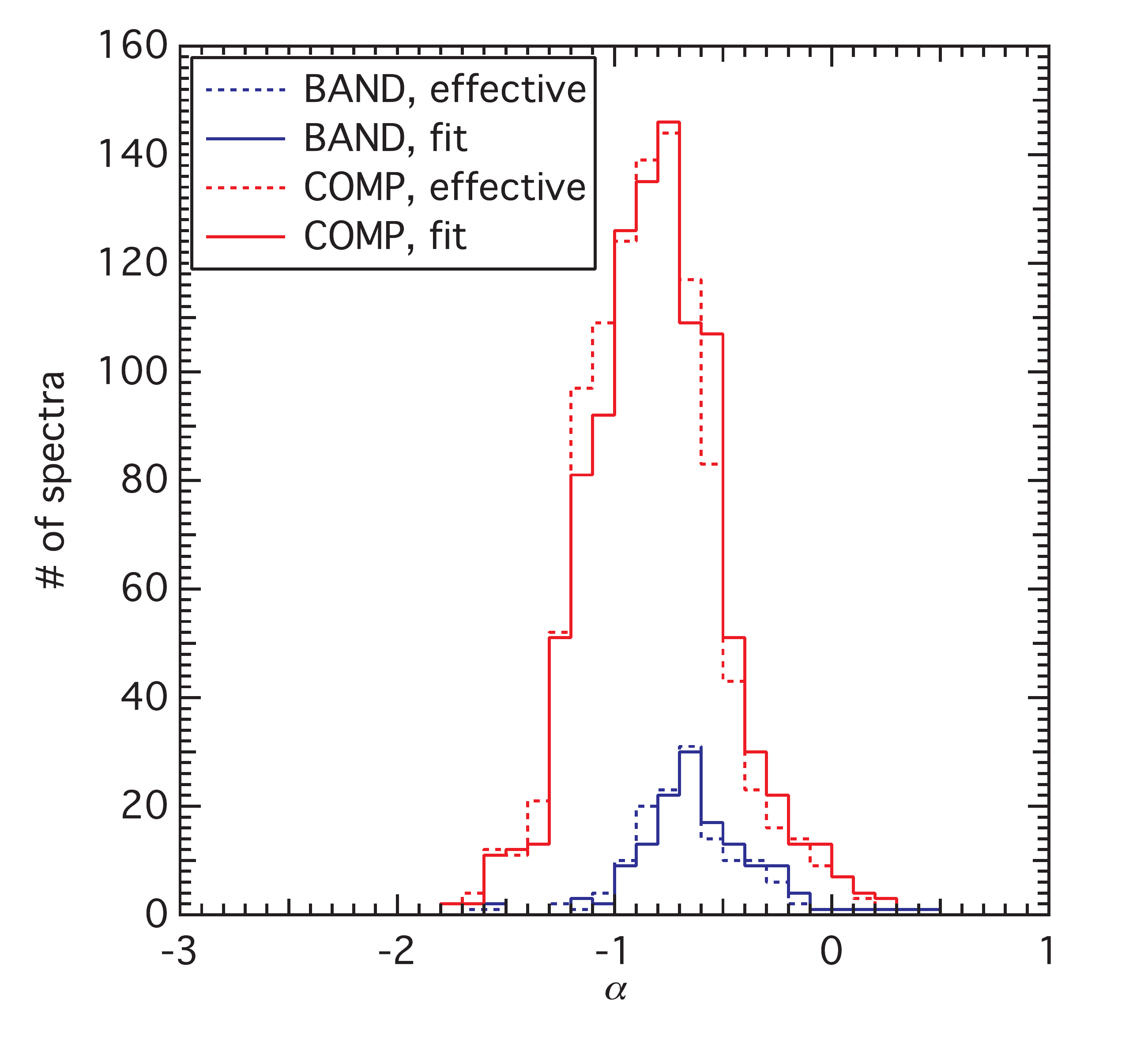}}
\caption{Histogram comparison of $\alpha_\text{eff}$ evaluated at 8~keV and $\alpha$. Blue and red histograms represent BAND and COMP, respectively. Solid lines are fit results, and dotted lines are effective $\alpha$.}
\label{fig:eff_hist}
\end{figure}

The cutoff power law, or the so-called Comptonized model (COMP), is a power-law model with a high-energy exponential cutoff. We note that when $\beta\to-\infty$, BAND reduces to COMP, as $E_\text{c}$ tends to infinity:
\begin{equation}
\label{eqn:comp}
f_\text{COMP}(E) = A \left(\frac{E}{100\text{ keV}}\right)^\alpha \exp\left[-\frac{(\alpha+2)E}{E_\text{p}}\right] \, ,
\end{equation}
where $A$ is the normalization factor at 100~keV in units of ph~s$^{-1}$~cm$^{-2}$~keV$^{-1}$, $\alpha$ is the power-law photon index, and $E_\text{p}$ is the peak energy in the $\nu F_\nu$ space in units of keV.

In the BATSE GRB spectral catalogs \citep{Pendleton94a,Preece00a,Kaneko06a,Goldstein13a}, the low-energy spectral index $\alpha$ of different models cannot be directly compared because they are asymptotic values but not actual slopes. They used an effective $\alpha$, $\alpha_\text{eff}$, computed at 25~keV (the BATSE detector lower limit). In the GBM GRB time-integrated spectral catalogs \citep{Goldstein12a,Gruber14a}, the fit values of $\alpha$ are directly adopted in their further analysis. Since GBM has a detector lower limit at 8~keV, the deviation from the asymptotic value \citep[Eqn.~C2,][]{Kaneko06a} is negligible (Figs.~\ref{fig:effective} and \ref{fig:eff_hist}), and here we follow the GBM GRB time-integrated spectral catalogs to use the best-fit values of $\alpha$ directly.

\subsection{The power law}

The power law (PL) is defined as
\begin{equation}
\label{eqn:pl}
f_\text{PL}(E) = A \left(\frac{E}{100\text{ keV}}\right)^\alpha \, ,
\end{equation}
where $A$ is the normalization factor at 100~keV in units of ph~s$^{-1}$~cm$^{-2}$~keV$^{-1}$, and $\alpha$ is the power-law photon index.

\subsection{Conditions to have a peak in the $\nu F_\nu$ space}

For all the aforementioned mathematical functions, the resulting spectrum has a peak in $\nu F_\nu$ space if and only if $\alpha > -2$ and $\beta < -2$. Since the Band function presumes $\alpha > \beta$, for the BAND fits with $\alpha \le -2$, the spectrum decreases monotonically, and for those with $\beta \ge -2$, the spectrum increases monotonically. For the SBPL fits with $\alpha \le -2$ or $\beta \ge -2$, $E_\text{p}$ is not calculated because Eqn.~(\ref{eqn:sbplep}) is invalid. Similarly, for the COMP fits with $\alpha \le -2$, $E_\text{p}$ is just a break and the spectrum decreases monotonically, and obviously not there for the PL model.

\section{Spectral analysis method}
\label{sect:ana}

The light curves are binned according to the procedure described in Sect.~\ref{sect:method}, resulting in a total of 1,802 time bins and $1,802 \times 4 = 7,208$ spectra. Time-resolved spectral analysis is performed using the official GBM spectral analysis software RMFIT v4.3BA\footnote{The public version of the RMFIT software package is available at \url{http://fermi.gsfc.nasa.gov/ssc/data/analysis/rmfit/}}, with effective area corrections applied to each pairs of NaI and BGO detectors. RMFIT employs a modified forward-folding technique based on the Levenberg-Marquardt algorithm. During the fitting process, the fitting models discussed in Sect.~\ref{sect:models} are converted into counts. These counts are then compared to the observed counts and RMFIT iterates itself until a best fit is found according to the chosen statistics for minimization.

In order to fold the model spectra into count space in the forward-folding process, detector response matrices (DRMs) are generated using the GBM response matrices v2.0. These DRMs contain information about the angular dependence of the detector efficiency, effective area of the detectors, partial energy deposition, energy dispersion, nonlinearity in the detectors, and atmospheric and spacecraft scattering of photons into the detectors. Therefore, they are functions of photon energies, angular dependence between spacecraft and the source, and angle between spacecraft orientation relative to the Earth. In order to account for the orientation change of the detectors relative to the burst direction because of the slew of the spacecraft, DRMs are generated for every 2\degr on the sky and grouped into RSP2 files for each burst. This means each DRM is weighted by the counts in the detectors for every 2\degr of spacecraft slew.

The chosen statistics for minimization in the fitting process is the so-called Castor C-Statistics (CSTAT). This is a modified statistical function based on the original Cash statistics \citep{Cash79a}. Since the background is Poissonian, the net count statistics is non-Gaussian, CSTAT is preferable over the traditional $\chi^2$ statistics. However, CSTAT does not provide a goodness-of-fit measure as $\chi^2$, because there is no standard probability distribution for the likelihood of CSTAT. As a result, test statistics must be calculated and compared to the resulting CSTAT values by simulating the fitting model a large number of times, which allows us to reject a model up to a certain confidence level. Theoretically, this should be done for each burst separately, but due to the infeasibility of generating large number of simulated spectra for all bursts, we adopt the values given in \citet{Gruber14a} for models (8.58 for PL vs. COMP, and 11.83 for COMP vs. BAND or SBPL) with various numbers of free parameters. These values, what we call the critical $\Delta\text{CSTAT}$ or $\Delta\text{CSTAT}_\text{crit}$, are listed in Table~1 of \citet{Gruber14a}.

\section{Results}
\label{sect:results}

\subsection{General statistics}

\begin{table}[!htbp]
\caption{Best-fit statistics for the BEST sample. For each sample the number of spectra $N$ and the percentage of the fraction of the spectra are given for each fitting model. ALL indicates parameter properties after combining the distributions (i.e., BAND + SBPL + COMP + PL).}
\label{tab:stat}
\centering
\def\arraystretch{1.5}
\begin{tabular}{ccc}
\hline\hline
Model & $N$ & percentage \\
\hline
BAND & 139 & 9.3\% \\
SBPL & 170 & 11.4\% \\
COMP & 1,030 & 69.1\% \\
PL & 152 & 10.2\% \\
ALL & 1,491 & - \\
\hline
\end{tabular}
\end{table}

\begin{table*}[!htbp]
\caption{The mean and median values of the best-fit parameters for the BEST sample. The mean values are computed by simply taking the averages of each parameter, and their errors are given by the standard deviations. The errors of the medians are given by the $1\sigma$ errors of each parameter by constructing the CDFs. ALL indicates parameter properties after combining the distributions (i.e., BAND + SBPL + COMP + PL). $^\text{a}$Due to the very different parameter behavior of PL, we give also the ALL without PL values which better reflect the statistics of the overall distribution of more complex models. $^\text{b}$The distributions of $E_\text{p}$ and $E_\text{b}$ are observed to be approximately log-normal, therefore we fit a log-normal distribution to each of the $E_\text{p}$ and $E_\text{b}$ populations, and reported the peak positions and 1$\sigma$ widths (in base-10 logarithmic of keV).}
\label{tab:para_stat}
\centering
\def\arraystretch{1.5}
\begin{tabular}{cccccccc}
\hline\hline
Parameter & Model & Mean & Median & Parameter & Model & \multicolumn{2}{c}{Peak$^\text{b}$}  \\
\hline
$\alpha$ & BAND & $-0.603\pm0.300$ & $-0.639^{+0.298}_{-0.205}$ & $\log_{10}(E_\text{p} / \text{keV})$ & BAND & \multicolumn{2}{c}{$\log_{10}(224.98) \pm 0.27$} \\
 & SBPL & $-0.763\pm0.362$ & $-0.741^{+0.241}_{-0.396}$ &  & SBPL & \multicolumn{2}{c}{$\log_{10}(165.79) \pm 0.40$} \\
 & COMP & $-0.802\pm0.312$ & $-0.810^{+0.287}_{-0.297}$ &  & COMP & \multicolumn{2}{c}{$\log_{10}(274.59) \pm 0.26$} \\
 & PL & $-1.674\pm0.169$ & $-1.648^{+0.147}_{-0.216}$ &  & - & \multicolumn{2}{c}{-}\\
 & ALL & $-0.867\pm0.413$ & $-0.823^{+0.304}_{-0.413}$ &  & ALL & \multicolumn{2}{c}{$\log_{10}(263.41) \pm 0.28$}\\
 & ALL w/o PL$^\text{a}$ & $-0.776\pm0.323$ & $-0.773^{+0.272}_{-0.320}$ &  & - & \multicolumn{2}{c}{-}\\
\hline
$\beta$ & BAND & $-2.214\pm0.272$ & $-2.183^{+0.224}_{-0.311}$ & $\log_{10}(E_\text{b} / \text{keV})$ & BAND & \multicolumn{2}{c}{$\log_{10}(129.71) \pm 0.22$}  \\
& SBPL & $-2.412\pm0.573$ & $-2.272^{+0.317}_{-0.573}$ &  & SBPL & \multicolumn{2}{c}{$\log_{10}(103.50) \pm 0.36$}\\
& ALL & $-2.323\pm0.472$ & $-2.217^{+0.262}_{-0.412}$ &  & ALL & \multicolumn{2}{c}{$\log_{10}(122.27) \pm 0.29$} \\
\hline
\end{tabular}
\end{table*}

\begin{figure*}
\resizebox{\hsize}{!}
{\includegraphics[width = 18 cm]{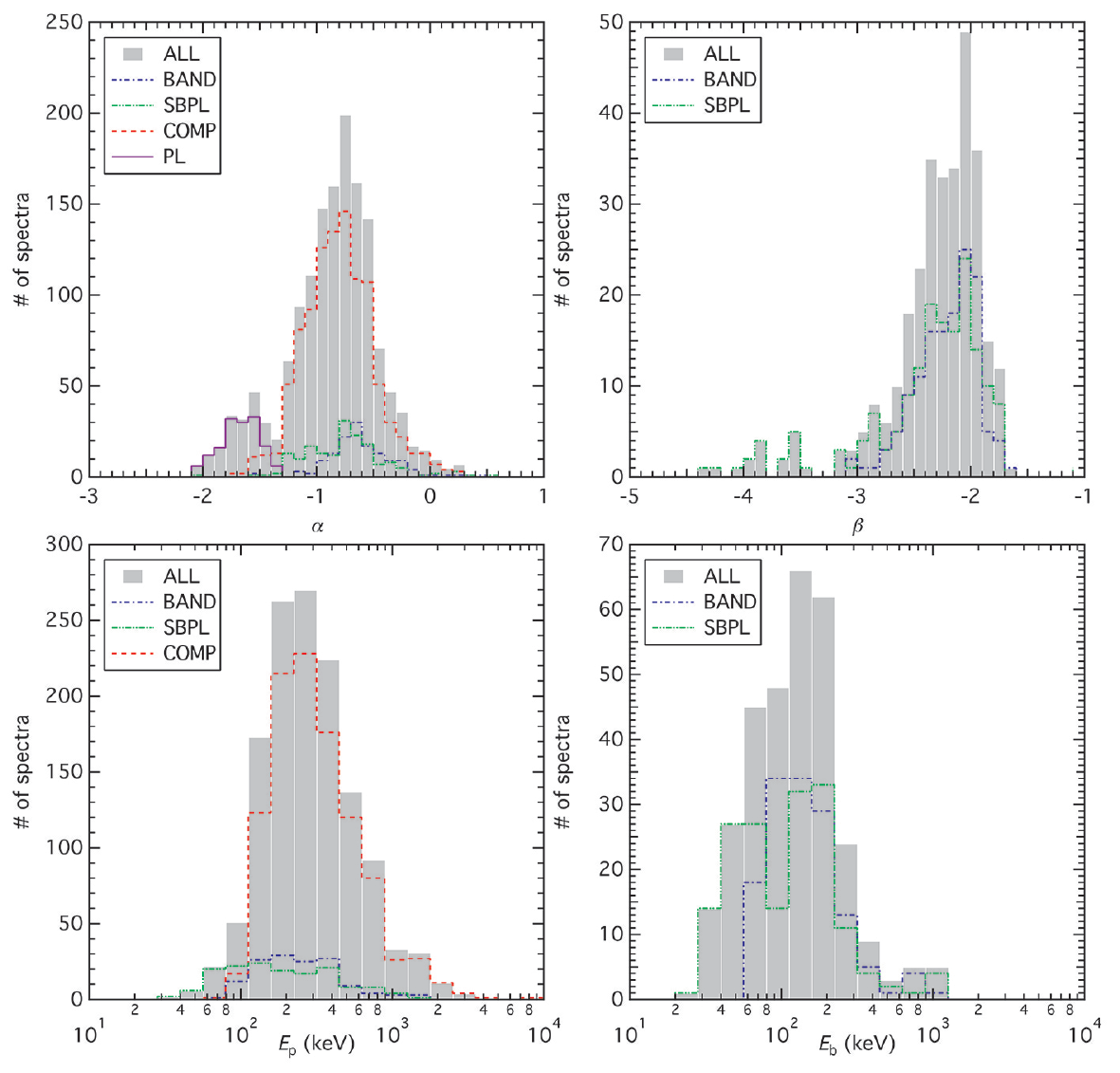}}
\caption{The distributions of the BEST sample spectral parameters. The BAND parameter populations are shown by the blue dash-dotted lines, the SBPL by the green dash-double dotted lines, the COMP by the red dashed lines, and the PL by the purple solid lines. The overall populations (ALL) are shown by the grey solid histograms. Top left panel: distributions of $\alpha$. Top right panel: distributions of $\beta$. Bottom left panel: distributions of $E_\text{p}$. Bottom right panel: distributions of $E_\text{b}$.}
\label{fig:BEST}
\end{figure*}

\begin{figure*}
\resizebox{\hsize}{!}
{\includegraphics[width = 18 cm]{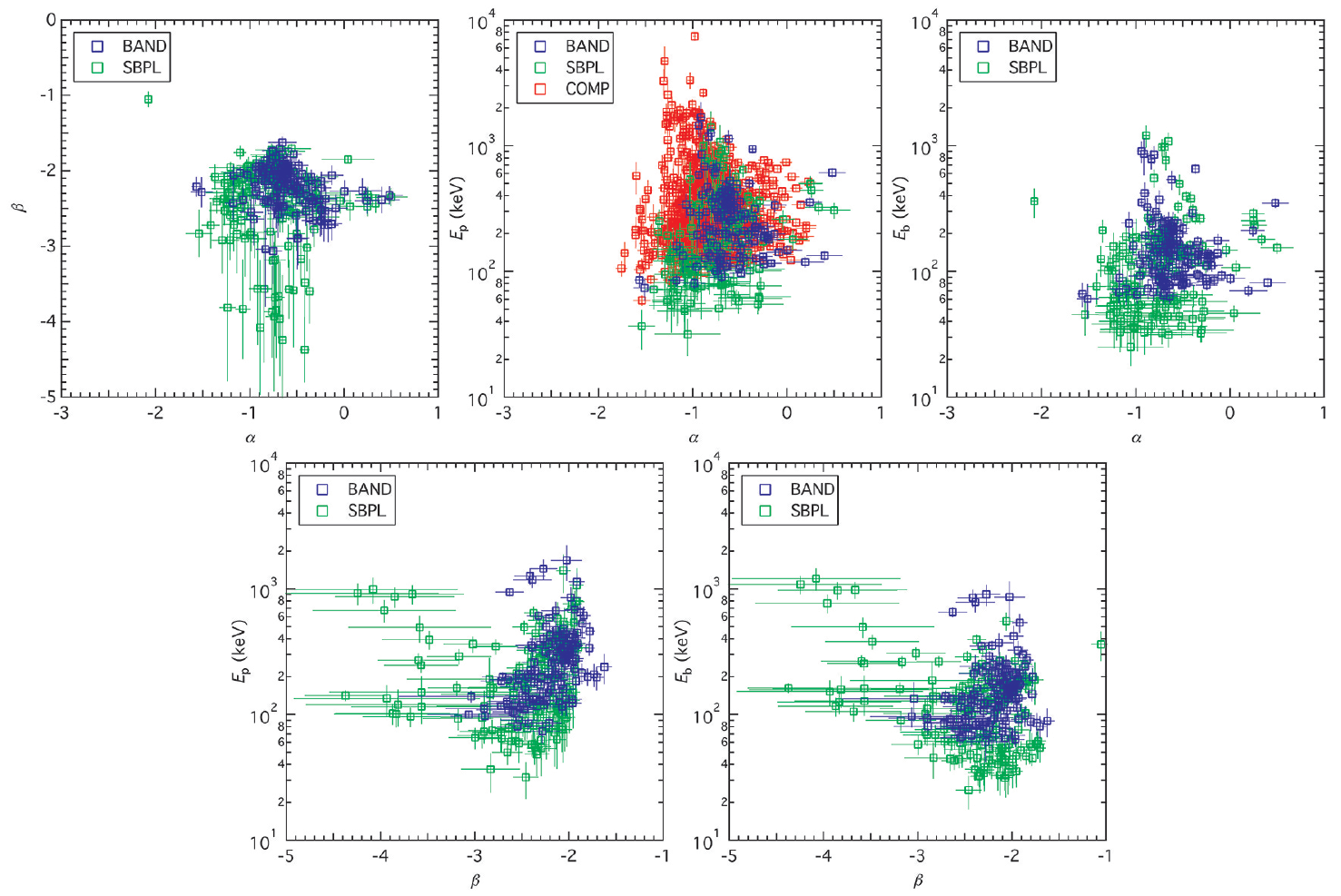}}
\caption{The scatter plots between the BEST sample spectral parameters. The blue, red, and green data points represent BAND, COMP, and SBPL fits, respectively. Top left panel: $\beta$ against $\alpha$. Top middle panel: $E_\text{p}$ against $\alpha$. Top right panel: $E_\text{b}$ against $\alpha$. Bottom left panel: $E_\text{p}$ against $\beta$. Bottom right panel: $E_\text{b}$ against $\beta$.}
\label{fig:para-para}
\end{figure*}

\begin{figure*}
\resizebox{\hsize}{!}
{\includegraphics[width = 18 cm]{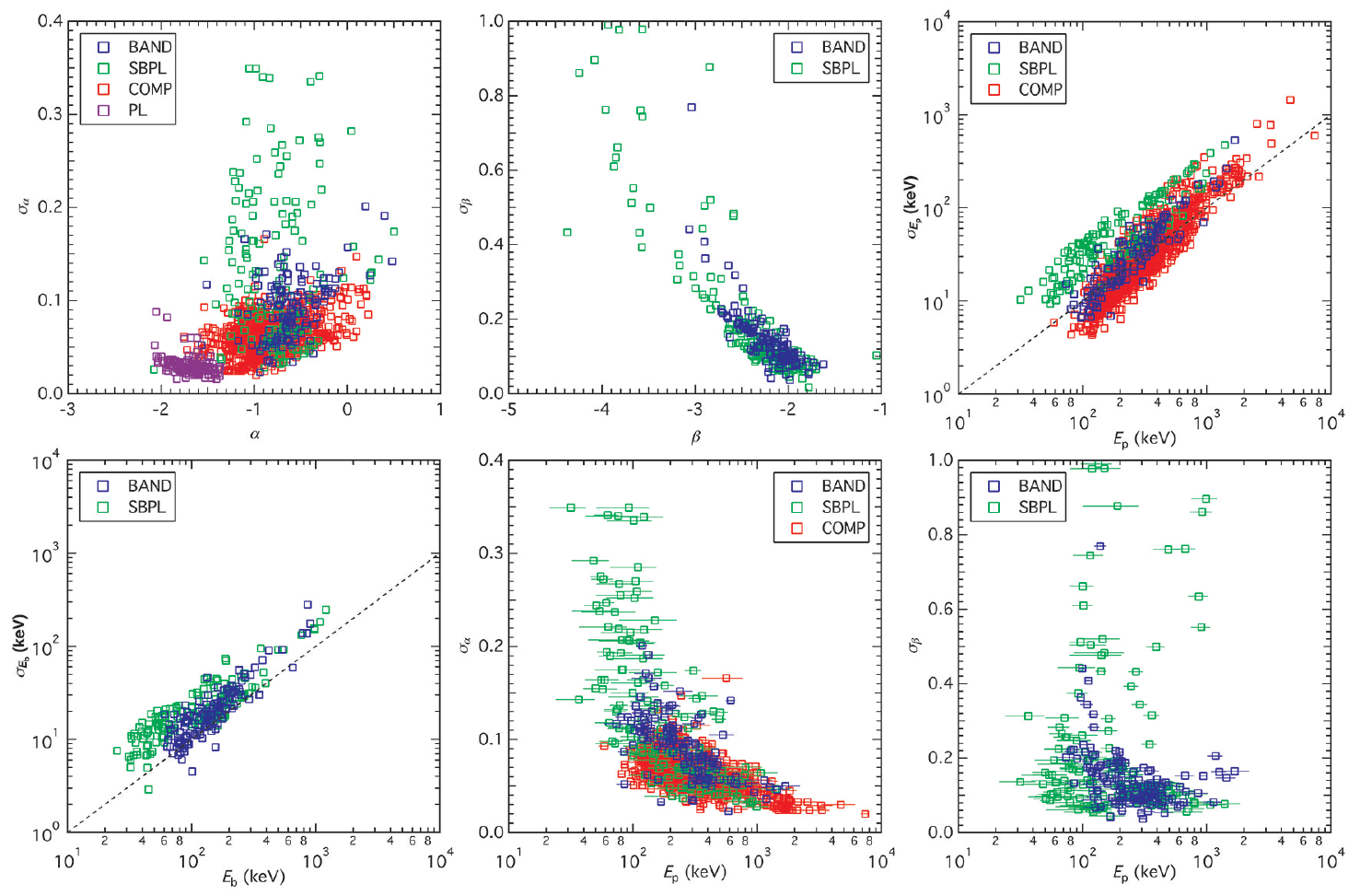}}
\caption{The scatter plots between the BEST sample spectral parameters and uncertainties. The blue, red, green, and purple data points represent BAND, COMP, SBPL, and PL fits, respectively. The dashed lines show $y = 0.1x$. Top left panel: $\sigma_\alpha$ against $\alpha$. Top middle panel: $\sigma_\beta$ against $\beta$. Top right panel: $\sigma_{E_\text{p}}$ against $E_\text{p}$. Bottom left panel: $\sigma_{E_\text{b}}$ against $E_\text{b}$. Bottom middle panel: $\sigma_\alpha$ against $E_\text{p}$. Bottom right panel: $\sigma_\beta$ against $E_\text{p}$.}
\label{fig:para-uncer}
\end{figure*}

\begin{figure*}
\resizebox{\hsize}{!}
{\includegraphics[width = 18 cm]{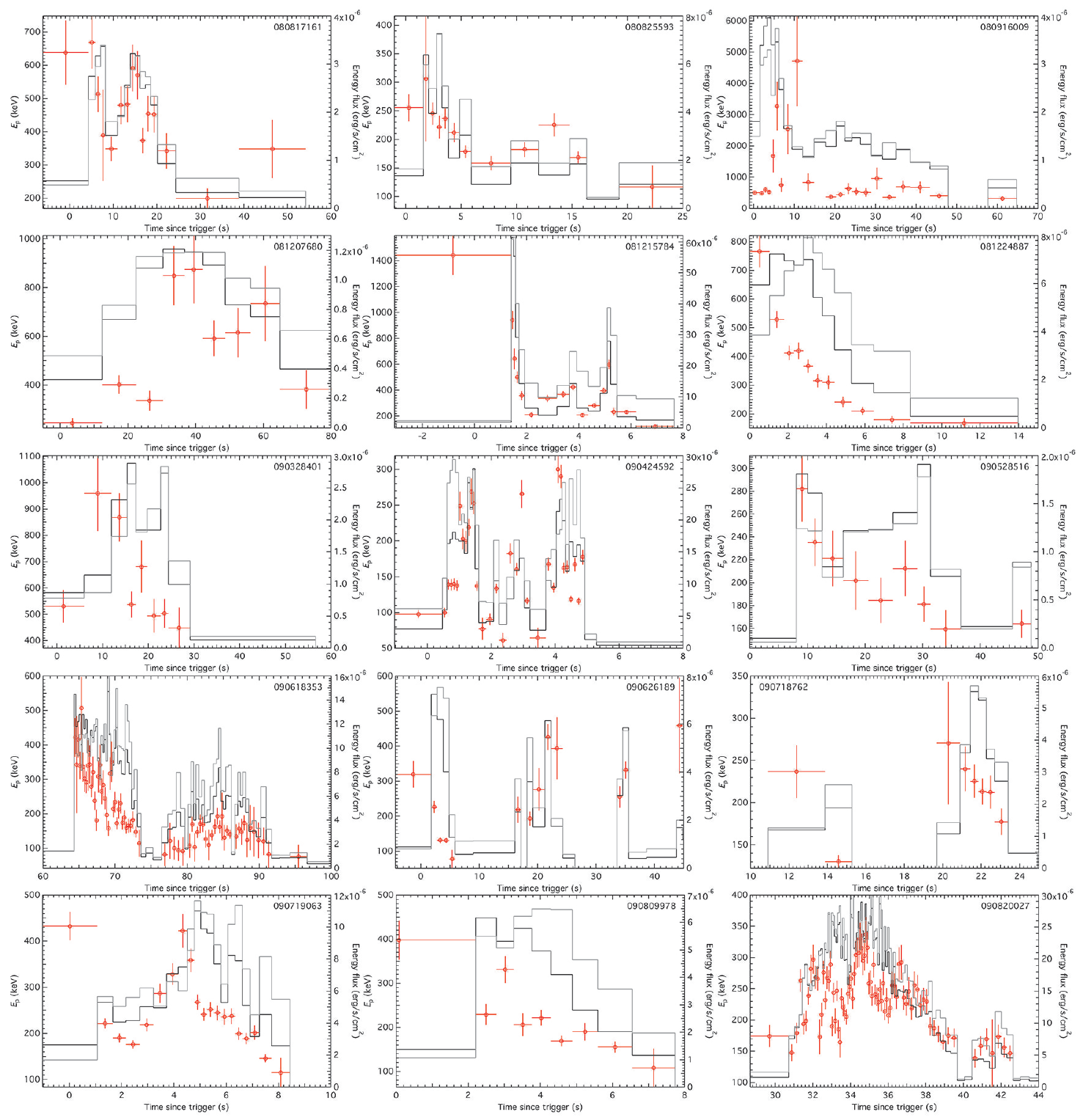}}
\caption{$E_\text{p}$ evolutions (red data points, left axis) of individual burst with the 10~keV - 1~ MeV energy flux (black histograms, right axis) and the 10~keV - 1~ MeV photon flux (grey histograms, arbitrary units) overlaid.}
\label{fig:Ep_evo1}
\end{figure*}

\begin{figure*}
\resizebox{\hsize}{!}
{\includegraphics[width = 18 cm]{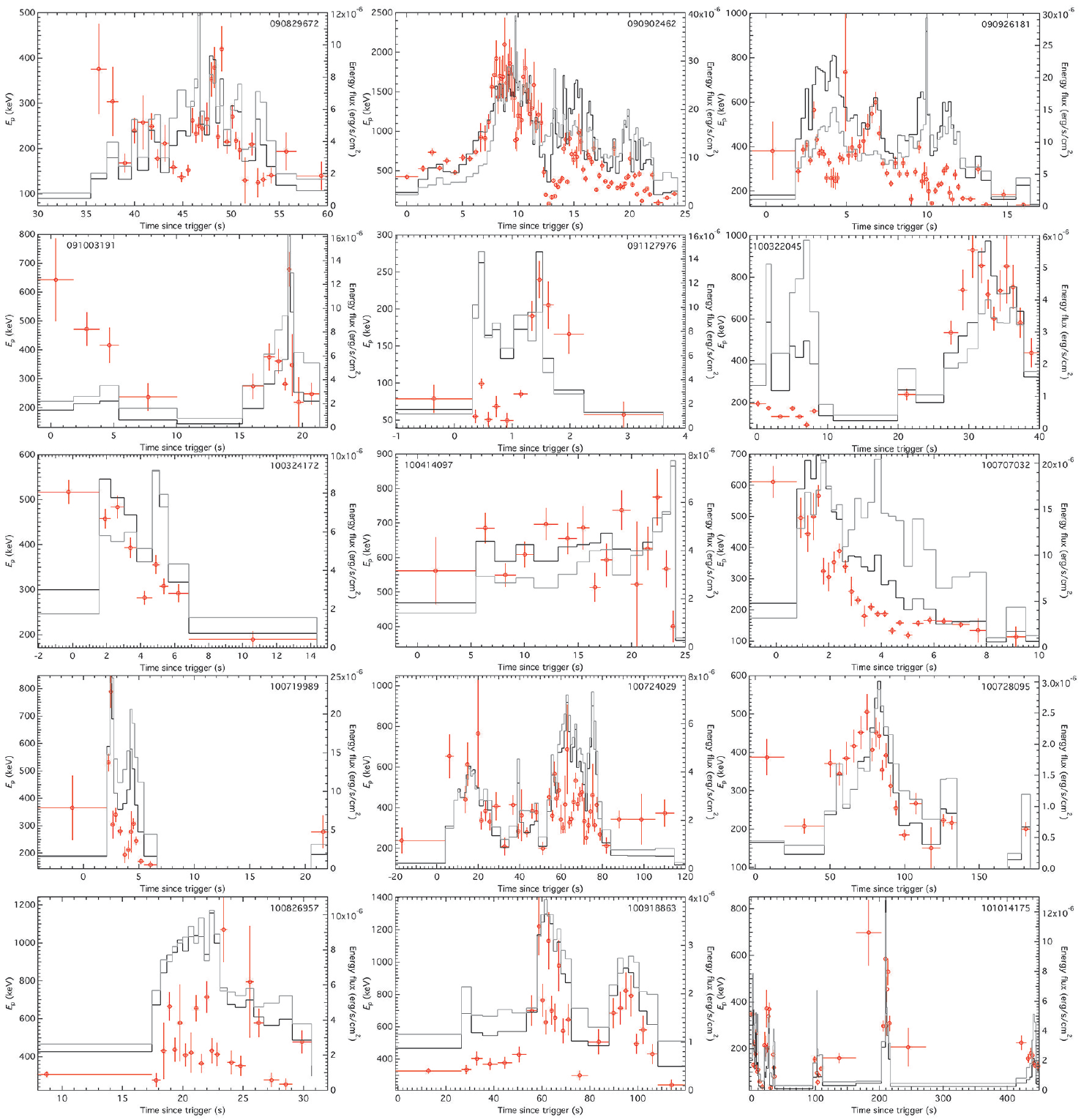}}
\caption{Same as Fig.~\ref{fig:Ep_evo1}.}
\label{fig:Ep_evo2}
\end{figure*}

\begin{figure*}
\resizebox{\hsize}{!}
{\includegraphics[width = 18 cm]{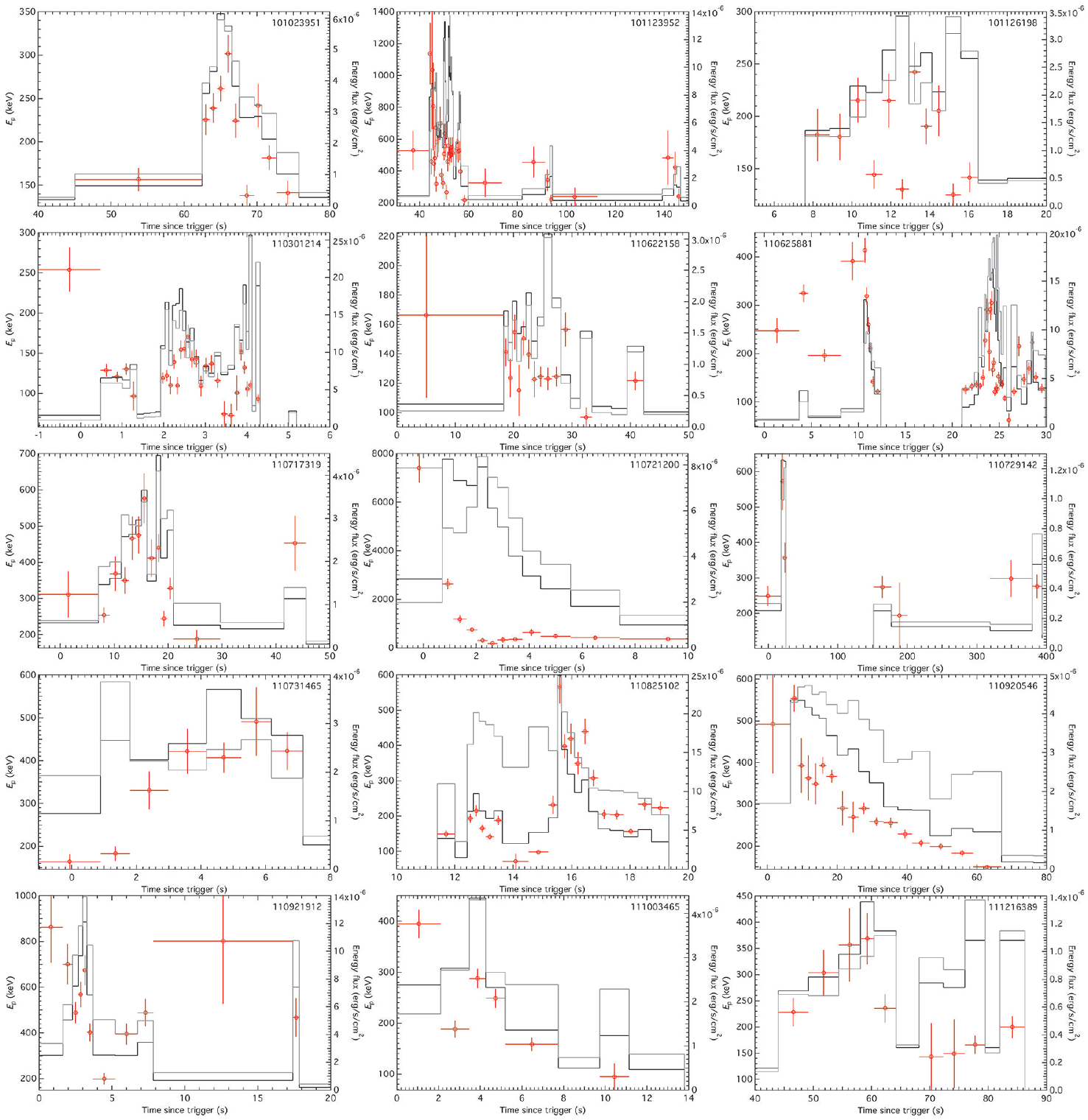}}
\caption{Same as Fig.~\ref{fig:Ep_evo1}.}
\label{fig:Ep_evo3}
\end{figure*}

\begin{figure*}
\resizebox{\hsize}{!}
{\includegraphics[width = 18 cm]{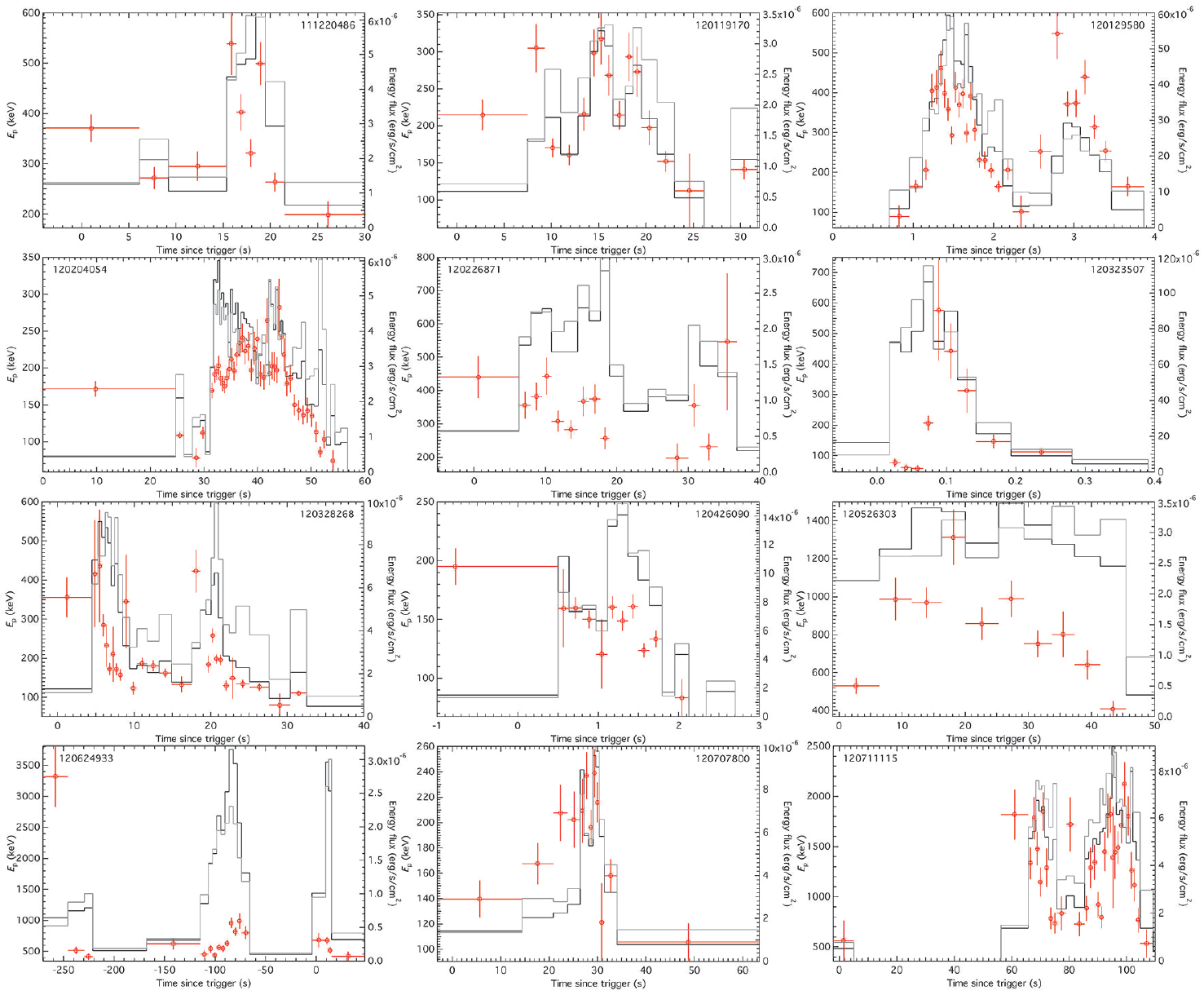}}
\caption{Same as Fig.~\ref{fig:Ep_evo1}.}
\label{fig:Ep_evo4}
\end{figure*}

We define the BEST model sample \citep[see][for example]{Goldstein12a,Gruber14a} by the following criteria: For each parameter $Q$ of a model, the relative error $\sigma_Q / Q \le 0.4$ except for all power-law indices; for models that have two power-law indices, the low-energy index error has to satisfy $\sigma_\alpha \le 0.4$, and the high-energy power-law index has to satisfy $\sigma_\beta \le 1.0$; for the single power law, the index error criterion is the same as $\alpha$'s; and the model has to have the lowest CSTAT after correcting the value by $\Delta\text{CSTAT}_\text{crit}$ (see Sect.~\ref{sect:ana}) comparing to other spectral model fits\footnote{In the first two GBM GRB time-integrated spectral catalogs, there was a definition of the GOOD sample. We do not include such a sample here in this catalog, since the GOOD criteria they used do not guarantee good fits. This is because the GOOD sample does not involve a goodness-of-fit criterion. We discuss this effect in Appendix~\ref{App:goodsample}.}. As a results, we are able to extract 1,491 BEST model fits out of the 1,802 spectra. The fit results of the BEST model of all spectra for all GRBs\footnote{In this paper, the names of the bursts are given according to the \textit{Fermi} GBM trigger designation that is assigned for each new trigger detected. The first 6 digits indicate the year, month, and day of the month, and the last 3 digits indicate the fraction of the day. For more details, please see the online \textit{Fermi} GBM burst catalog at \url{http://heasarc.gsfc.nasa.gov/W3Browse/fermi/fermigbrst.html}} are listed in Table~\ref{tab:bigtable}.

We note that BAND's $E_\text{b}$ and SBPL's $E_\text{p}$ are computed instead of fit parameters. Therefore, we compute $\sigma_{E_\text{b}}$ of BAND and $\sigma_{E_\text{p}}$ of SBPL by performing Monte-Carlo simulations using the errors of the best-fitting model parameters. We randomly draw new values of the model parameters from a uniform probability function sharing the same $1\sigma$ width. This process is repeated to generate 10,000 realizations, and a cumulative distribution function (CDF) is constructed. The errors are then obtained by taking the $1\sigma$ width of the resulting CDFs. This procedure generates the most conservative error values because the uniform probability function has the largest standard deviation.

The fit statistics for the BEST sample are listed in Table~\ref{tab:stat}. It can be seen that COMP has the largest fraction of BEST fits (69.1\%), SBPL and PL have 11.4\% and 10.2\% respectively, and BAND gives the lowest fraction, only 9.3\%. However, we note that these resulting statistics do not necessarily imply that the Comptonized model is generally favored over the Band function. \citet{Kaneko06a} and \citet{Goldstein12a} showed that there appeared to be a strong correlation between the S/N and the complexity of the BEST model. Therefore, we cannot rule out the possibility that this observed preference is due to poor count statistics at the high energies.

The mean and median values of the parameter distributions for the BEST sample are shown in Table~\ref{tab:para_stat}. The "Mean" columns show the average value of each parameter distribution and their errors are given by the standard deviations. The "Median" columns show the median and $1\sigma$ errors of each parameter by constructing the CDFs. For the approximately log-normally distributed $E_\text{p}$ and $E_\text{b}$ populations, log-normal distributions are fit to each population and the peak and 1$\sigma$ widths in base-10 logarithmic space are reported.

In Fig.~\ref{fig:BEST} we show the distributions of the BEST sample best-fit parameters. The top left panel shows the BEST distributions of the low-energy power-law index $\alpha$. It can be seen that there are two peaks in the ALL population. The peak at $\alpha \approx -0.7$, excluding those values from PL, is dominated by the COMP model. It can be seen that the population of SBPL's $\alpha$ is slightly softer than that of the BAND's and COMP's and also shows a larger spread. As discussed above, this effect is not due to the detector's lower limit because the histogram's bin width is wider than the deviation from the asymptotic limit. The $\alpha$ of PL fits are significantly softer than that of other models, with no $\alpha > -1.3$. The distinct behavior of PL to the other fit functions is evident.

The top right panel shows the BEST distributions of the high-energy power-law index $\beta$. It can be seen that the BAND's $\beta$ becomes more concentrated between $-3.1$ and $-1.6$, while the SBPL's $\beta$ extend to much steeper values of about $-4.4$.\footnote{BAND with $\beta = -4.4$ effectively mimics COMP, while SBPL with $\beta = -4.4$ does not. This is because the mathematical definitions of the curvatures of SBPL and BAND are different.} The peak of the populations is at $\beta \approx -2.1$. As a result, 21\% of the overall population of $\beta \ge -2$ (no peak in the $\nu F_\nu$ space).

The bottom left panel shows the BEST distributions for the $\nu F_\nu$ peak energy $E_\text{p}$. It can be seen that the $E_\text{p}$ population of COMP dominates the overall distribution, and that the COMP population extends to higher energies (up to about 5~MeV) than the BAND and SBPL populations, which instead extend to lower energies (down to about 20~keV). We do not find any spectrum with very large $E_\text{p}$, with only 4.8\% of the overall population of $E_\text{p} \ge 1$~MeV.

The bottom right panel shows the distributions for the break energy $E_\text{b}$. It can be seen that the $E_\text{b}$ population of BAND has a clear peak at $E_\text{b} \approx 130$~keV, while the $E_\text{b}$ population of SBPL has a broad distribution (from 40~keV to 300~keV).

These general statistics suggests that when performing spectral analysis of GRBs, one should not assume a Band spectrum \citep[e.g.,][]{Giblin99a,Gonzalez12a,Sacahui13a}. Instead, one should always try different fit functions and compare the fit statistics to find the best description to the data. Similar statistical behaviors are also observed in the time-integrated spectral catalogs \citep{Goldstein12a,Gruber14a}.

\subsection{The parameter-parameter scatter plots}
\label{para-para}

Figure~\ref{fig:para-para} shows the scatter plots between the best-fit parameters of the BEST sample. The top left panel shows the plot of $\beta$ against $\alpha$. Trends can neither be found between $\alpha$ and $\beta$ for individual models nor the overall population as a whole. It can be seen that SBPL's $\beta$ population extend to steeper values and have larger error bars in the same range of values of $\alpha$. The larger error for steeper $\beta$ shows that the SBPL tends to mimic a COMP spectrum, in which $\beta$ is poorly constrained due to less photon statistics at the higher energies.

The top middle panel shows the plot of $E_\text{p}$ against $\alpha$. Trends can neither be found between $\alpha$ and $E_\text{p}$ for individual models nor the overall population as a whole. It can be seen that while the data points seem to occupy the same region, SBPL's $E_\text{p}$ extends to lower energies and COMP's $E_\text{p}$ extends to higher energies, for similar range of values for $\alpha$.

The top right panel shows the plot of $E_\text{b}$ against $\alpha$. Similar to the plot of $E_\text{p}$, no trends can be found for $E_\text{b}$, and SBPL's $E_\text{b}$ extends to lower energies for similar range of values of $\alpha$. This is because, according to Eqns.~(\ref{eqn:bandeb}) and (\ref{eqn:sbplep}), $E_\text{p}$ is proportional to $E_\text{b}$.

The bottom left panel shows the plot of $E_\text{p}$ against $\beta$, and the bottom right panel shows the plot of $E_\text{b}$ against $\beta$. Since $E_\text{p}$ is proportional to $E_\text{b}$, the two plots show similar behaviors. A slight trend may exist between $E_\text{p}$ against $\beta$ in the population of BAND: steeper $\beta$ tends to have lower $E_\text{p}$. However, this trend is not seen in the population of SBPL.

These plots show that the SBPL produces larger uncertainties for steeper $\beta$, and has difficulties to constrain the high-energy power-law behavior in comparison to the Band function.

\subsection{The parameter-uncertainty scatter plots}
\label{para-uncer}

Figure~\ref{fig:para-uncer} shows the scatter plots between the best-fit parameters and uncertainties of the parameters of the BEST sample. The top left panel shows the plot of $\sigma_\alpha$ against $\alpha$. It is seen that the SBPL gives the most scatter and large errors (extend to almost $\sigma_\alpha = 0.4$), while other models give relatively small errors of $\sigma_\alpha < 0.2$. The PL gives the smallest $\sigma_\alpha \le 0.1$. A clear trend for $\sigma_\alpha$ can be seen: $\sigma_\alpha$ tends to be larger when $\alpha$ increases (i.e., becomes harder).

The top middle panel shows the plot of $\sigma_\beta$ against $\beta$. A clear trend is observed that $\sigma_\beta$ becomes larger when $\beta$ decreases (i.e., becomes softer/steeper). The trend is indeed expected because the high-energy power-law slope becomes less constrained when the BAND or SBPL mimics a COMP model, i.e., when there are less photon statistics at the high energies which leads to a cutoff behavior.

The top right panel shows the plot of $\sigma_{E_\text{p}}$ against $E_\text{p}$. It is observed that $\sigma_{E_\text{p}}$ of SBPL is systematically larger than that of BAND and COMP for the same value of $E_\text{p}$. The values of $\sigma_{E_\text{p}}$ for BAND and COMP also tend to lie above the dashed line, implying that $\sigma_{E_\text{p}}$ becomes larger when $E_\text{p}$ increases. We note that $E \approx 900$~keV is the upper energy boundary of the NaIs, so that there are only data contributed by the BGOs beyond this limit, providing less photon statistics and thus increases the uncertainty in determining the spectral peak position.

The bottom left panel shows the plot of $\sigma_{E_\text{b}}$ against $E_\text{b}$. Comparing to the peak energies, $\sigma_{E_\text{b}}$ of the break energies  $E_\text{b}$ have similar trends for both the BAND and SBPL fits. The errors lie systematically above the dashed line for both models.

It is also of interest to investigate how the position of the spectral peak affects the uncertainties in the spectral slopes. The bottom middle panel shows the plot of $\sigma_\alpha$ against $E_\text{p}$. A clear trend is observed that the low-energy power-law slope becomes more uncertain when the spectrum peaks at lower energies. This is because the low-energy spectral slope is determined by the photon statistics below the peak energy. When the peak energy is smaller, there are relatively fewer data points to constrain the value of the low-energy power-law slope. It is also observed that for the same value of $E_\text{p}$, $\sigma_\alpha$ tends to be larger for the SBPL fits than that for the BAND or COMP fits.

The bottom right panel shows the plot of $\sigma_\beta$ against $E_\text{p}$. A trend is observed that higher values of $E_\text{p}$ tend to produce smaller $\sigma_\beta$, which is weaker in comparison to the plot of $\sigma_\alpha$ against $E_\text{p}$. This shows that the high-energy power-law slope is not as strongly coupled to the peak position as the low-energy power-law slope.

These plots again show that the smoothly broken power-law model produces the highest degree of uncertainties in the best-fit parameters. This is not limited in the high-energy power-law index $\beta$, as shown in Fig.~\ref{fig:para-para}. Figure~\ref{fig:para-uncer} shows that SBPL's peak position significantly affects the uncertainties of both power-law indices, more so than the other models. The slight offsets of the best-fit parameters from different fit functions are expected because they have intrinsically different parametrical formulae. In general, we observe good consistency in the parameter space occupation, indicating that the minima in the parameter spaces are well defined and our results are statistically reliable.

\subsection{$E_\text{p}$ evolution}

Time-resolved spectral analysis of GRBs has shown that there are two different kinds of $E_\text{p}$ evolutionary trends \citep[e.g.,][]{Ford95a}, namely the intensity tracking and the hard-to-soft behavior. Intensity tracking bursts show evidence that the values of $E_\text{p}$ follow similar trends of the intensity (either photon flux or energy flux) in their light curves. Hard-to-soft bursts show evidence that $E_\text{p}$ decays (in general) monotonically with time.

We compute the Spearman's Rank Correlation Coefficient $\rho$ \citep{Spearman04a} between $E_\text{p}$ and (1) the 1~keV - 1~MeV photon flux,  $\rho_\text{ph}$, (2) the 1~keV - 1~MeV energy flux, $\rho_\text{en}$, and (3) the time, $\rho_\text{t}$. A positive value indicates a positive correlation, a negative value indicates a negative correlation, and a value of zero means no correlation. The process is repeated for different confidence levels of 90\%, 95\%, and 99\%. We note that the confidence levels are \textit{not} the probabilities to find $\rho$ within the confidence intervals. They are the \textit{ratios} of finding the real $\rho$ within the confidence intervals to the total number of repeated analysis. For example, the 99\% confidence interval of $\rho$ denotes that if the spectral analysis is repeated a large number of times, we will find on average, in 99 out of 100 times, that the real $\rho$ lies within the 99\% confidence interval. However, we will never know if we have picked the lucky ones, because we have no way to know the actual value of $\rho$. Therefore, the confidence level of a confidence interval provides a sense of how often a correlation is expected to be found.

First, we distinguish the $E_\text{p}$ evolutionary trends by machine. For each confidence level of the $\rho$'s, we check the following logical criteria:\footnote{We iterated the machine-based decision process for many different logical criteria, and found that the stated criteria provide a fairly robust determination of the trends comparing to human decisions. See main text and Table~\ref{tab:trends}.} (1) if $\rho_\text{ph}$ \textit{or} $\rho_\text{en} > 0.5$, \textit{and} it is not consistent with zero within the confidence interval; \textit{and} $\rho_\text{t} \ge -0.5$ \textit{or} it is consistent with zero within the confidence interval, then we define the trend as intensity tracking ("in.track."); (2) if $\rho_\text{ph} \le 0.5$ \textit{and} $\rho_\text{en} \le 0.5$, \textit{or} they are consistent with zero within their confidence intervals; \textit{and} $\rho_\text{t} < -0.5$ \text{and} it is not consistent with zero within the confidence interval, then we define the trend as hard-to-soft ("h.t.s."); (3) everything else are defined as undetermined ("undeter."). The values and confidence intervals of the $\rho$'s, and the machine-decided kinds of trends are listed in Cols.~(3) - (14) of Table~\ref{tab:trends}.

Then, we distinguish the $E_\text{p}$ evolutionary trends by human eyes (Col.~15 of Table~\ref{tab:trends}). We plot the $E_\text{p}$ evolutions (red data points, left axis) in Figs.~\ref{fig:Ep_evo1} - \ref{fig:Ep_evo4}, with the 10~keV - 1~ MeV energy flux (black histograms, right axis) and the 10~keV - 1~ MeV photon flux (grey histograms, arbitrary units) light curves overlaid. We note that we only plot and compare the 57 bursts with $E_\text{p}$ in at least 6 time bins or more. We find that the machine-based decision process is quite robust, in that only 2 bursts (3.5\%) are mis-attributed to the opposite kind ("h.t.s." vs. "in.track."), namely GRB 100719989 (Fig.~\ref{fig:Ep_evo2}) and GRB 111216389 (Fig.~\ref{fig:Ep_evo3}). The brightness of the first peak relative to the second peak of GRB 100719989 mimics a trend that $E_\text{p}$ is decaying with time. In contrast, a human would identify its intensity tracking nature by noticing the low $E_\text{p}$ in the first time bin and the small rise of $E_\text{p}$ values during the second peak. The case of GRB 111216389 is similar in that the relatively higher value but intensity tracking $E_\text{p}$ during the first peak to the second peak contributed to a small excess in $\rho_\text{t}$.

There are 12 GRBs (21\%) which show a mix of the two kinds of trends. Some of these bursts are identified by the computer as either one of the two kinds, or as undetermined. Two of them are especially worth mentioning: GRB 090618353 (Fig.~\ref{fig:Ep_evo1}) and GRB 091003191 (Fig.~\ref{fig:Ep_evo2}). They both show an initial hard-to-soft evolution followed by a later intensity tracking behavior, where the computer labeled them as undetermined. The other 10 bursts show a general hard-to-soft decay of $E_\text{p}$ where the values in between seem to follow the intensity profile. \citet{Lu12a} have shown that intrinsic hard-to-soft evolutions of distinct pulses can overlap and produce such a "h.t.s.+in.track." behavior. They claimed that both "h.t.s." and "in.track." behavior could be intrinsic to a burst or a pulse, which is consistent with our findings that many single pulsed bursts show pure intensity tracking behavior. We also find that the intensity tracking behavior of $E_\text{p}$ with the energy flux is more prominent than that with the photon flux in all of the intensity tracking bursts.

A few more bursts are worth of mentioning. GRB 100707032 (Fig.~\ref{fig:Ep_evo2}), GRB 110721200, and GRB 110920546 (both Fig.~\ref{fig:Ep_evo3}) are single pulsed, fast-rise-exponential-decay (FRED) bursts. All of them show pure hard-to-soft behavior. Since the $E_\text{p}$ evolutions and intensity profiles of these FRED bursts are very similar, $\rho_\text{en}$ and $\rho_\text{ph} \approx -\rho_\text{t} \gtrsim 0.5$, thus the computer cannot determine their evolutionary trends.

In short, we emphasize that even though the process of distinguishing "h.t.s." and "in.track." bursts can be done automatically, the existence of "h.t.s.+in.track." and FRED bursts can be ambiguous to computers. We strongly encourage checking by human eyes after any automated detection process of $E_\text{p}$ evolutionary trends.

\subsection{Search for blackbody emission}
\label{subsect:blackbody}

Many studies have reported evidence for thermal components with $kT \sim 10$~keV in various GRBs \citep[e.g.,][]{Meszaros02a,Ryde05a,Guiriec11a,Axelsson12a,Guiriec13a,Burgess14a,Burgess14b,Guiriec15a,Guiriec15b,Peer15b,Iyyani15a}. Therefore, adding a blackbody component (i.e., a Planck function) to the fit function is a natural way to explore the data in this time-resolved catalog. The blackbody model (BB) is defined as:
\begin{equation}
\label{eqn:bb}
f_\text{BB}(E) = A \left[\frac{(E/1\text{ keV})^2}{\exp(E/kT)-1}\right] \, ,
\end{equation}
where $A$ is the normalization factor at 1~keV and $kT$ is the blackbody temperature in units of keV.

We find that except for the single power law, in most of the cases ($\gtrsim 90$\%) it is not possible to obtain converged fits when the blackbody is added to other fitting models (i.e., BAND, COMP, and SBPL). However, we note that the ability of a model to fit the data depends also on the count statistics. \citet{Abdo09a} performed a joint GBM-LAT analysis to GRB 090902B (GBM trigger \#090902462, see discussion below) that they can fit a BAND plus PL model to the burst, which is expected because there are more statistics to constrain more parameters. Our results in this paper indicates the difficulty of fitting a model with 5 or more parameters to the GBM data alone using the S/N $=30$ criterion.

The power law plus blackbody model (PLBB) is defined as
\begin{equation}
\label{eqn:plbb}
f_\text{PLBB}(E) = A_\text{PL} \left(\frac{E}{100\text{ keV}}\right)^\alpha + A_\text{BB} \left[\frac{(E/1\text{ keV})^2}{\exp(E/kT)-1}\right] \, ,
\end{equation}
where $A_\text{PL}$ and $A_\text{BB}$ are the normalization factors for the power-law and blackbody component, respectively.

Since PLBB is a not a nested model, it is necessary to perform $\Delta\text{CSTAT}_\text{crit}$ simulations for every pair of competing models, instead of just counting the number of free parameters \citep[see][]{Gruber14a}. However, doing a large number of simulations for every spectrum is obviously impractical. We therefore first identify plausibly significant PLBB spectra by using the same $\Delta\text{CSTAT}_\text{crit}$ criteria for a 4-parameters model. Then we generate 10,000 realisations for the identified time intervals for every burst in this subsample and obtain the $\Delta\text{CSTAT}_\text{crit}$ for each burst. Then we compare the $\Delta\text{CSTAT}$ between the BEST model and the PLBB model for each spectrum, i.e., $\Delta\text{CSTAT} = \text{CSTAT(BEST)} - \text{CSTAT(PLBB)}$.

\begin{table}
\caption{The 4 bursts with $N$ number of PLBB-identified spectra, and their respective critical $\Delta\text{CSTAT}$ values.}
\label{tab:PLBBs}
\centering
\def\arraystretch{1.5}
\begin{tabular}{ccc}
\hline\hline
GRB name & $N$ & $\Delta\text{CSTAT}_\text{crit}$ \\
\hline
090618353 & 2 & 19.55 \\
090902462 & 32 & 32.75 \\
110622158 & 2 & 12.32 \\
110920546 & 6 & 148.37 \\
\hline
\end{tabular}
\end{table}

As a matter of fact, 56 plausibly significant PLBB spectra are identified among 16 bursts, in which 14 bursts have only 1 plausible spectrum identified. Since a blackbody component is likely to be present in multiple spectra within a burst if it is real, we drop these 14 bursts and concentrate on the remaining 4 bursts (42 spectra in total) with multiple PLBB-identified spectra. These bursts are listed in Table~\ref{tab:PLBBs}, and their simulated $\Delta\text{CSTAT}_\text{crit}$ values are also given.

We find that the spectra of the bursts listed in Table~\ref{tab:PLBBs} have $\Delta\text{CSTAT} > \Delta\text{CSTAT}_\text{crit}$, except for GRB 110920546. These 36 PLBB spectral parameters are listed in Table~\ref{tab:blackbodies}. We note that the 4 PLBB spectra from GRB 090618353 and GRB 110622158 have values of $kT \sim 20$~keV, while the 32 spectra from GRB 090902B show $kT \sim 200$~keV. In GRB 090902B, \citet{Abdo09a} identified an extra power-law component on top of Band functions with hard values of $E_\text{p}$ using wider time bins and joint GBM-LAT data. Using only the GBM data, we find that most of the BAND plus PL fits of our more resolved time bins in this time interval are either poorly constrained, unconstrained, or even not converged; but, interestingly, our values of the PL indices are very similar to theirs ($\alpha \approx -1.8$). This indicates the ability of a model to fit data depends on (1) how many free parameters (in this case, BB vs. BAND), and (2) the count statistics (GBM alone vs. GBM-LAT). We also note that \citet{Peer12a} used a thermal plus non-thermal theoretical model to apply to the spectra of this burst, in which they claimed that the data are consistent with such a hybrid emission model.

We note that the $\Delta\text{CSTAT}_\text{crit}$ can vary much across different bursts. Recently, \citet{Burgess15a} showed that it is very plausible to get false positive for an extra blackbody component in time-integrated spectra due to severe spectral evolution. Therefore, we recommend researchers to perform independent simulations on time-resolved spectra for different bursts in order to reduce the chance of false positives.

\subsection{Comparison to time-integrated results}
\label{subsect:compare}

\begin{figure*}
\resizebox{\hsize}{!}
{\includegraphics[width = 18 cm]{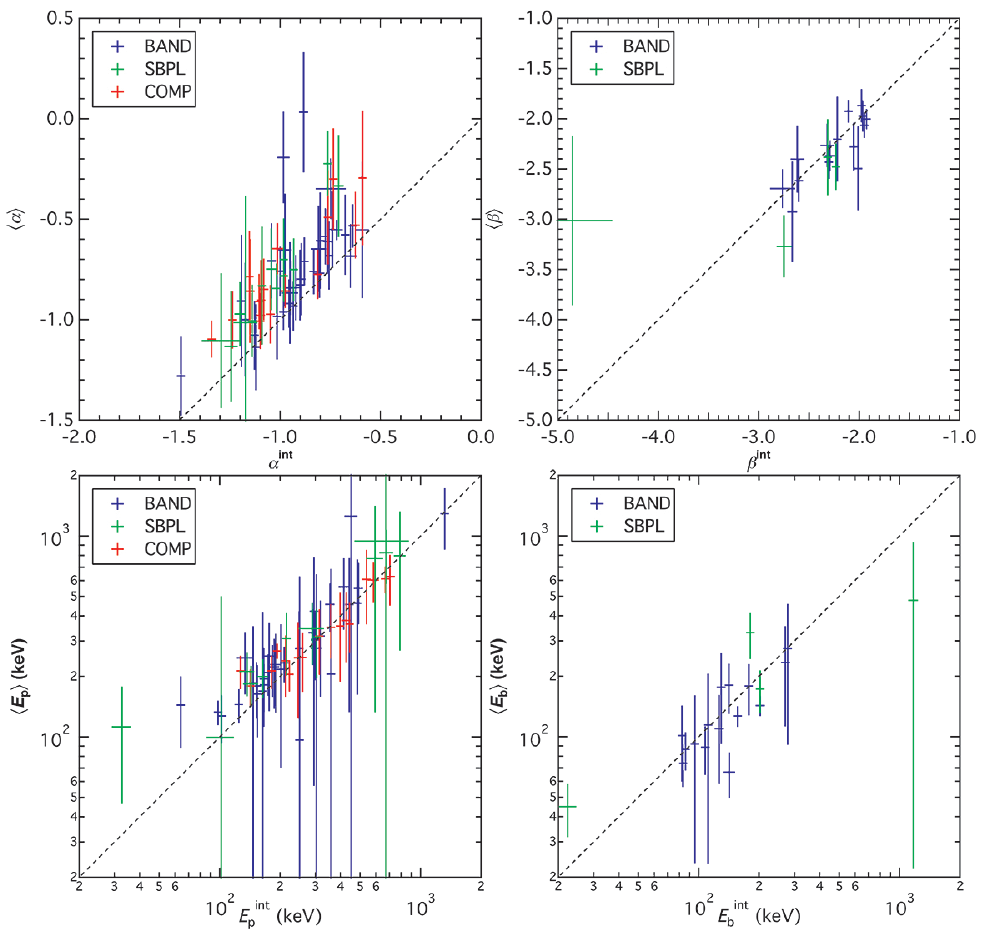}}
\caption{Comparisons between the averaged time-resolved BEST parameters (vertical axis) and the time-integrated BEST parameters (horizontal axis) for each burst in this catalog. The errors of the averaged time-resolved parameters are given by their standard deviations within the burst. Blue, green, and red data points represent the time-integrated BEST model is BAND, SBPL , and COMP, respectively. The diagonal dashed lines show $x=y$. Only averaged time-resolved parameters determined in 5 or more time bins are included.}
\label{fig:averaged}
\end{figure*}

This catalog made the comparison between time-resolved and time-integrated fit parameters of a large sample of GRBs possible. Here we compare our time-resolved results with the time-integrated results from \citet{Gruber14a}.

Comparing Fig.~\ref{fig:BEST} to Figs.~3, 4, 6, \& 7 in \citet{Gruber14a}, no significant difference between the overall parameter populations is observed. However, using the same burst sample, \citet{Yu15b} observed a significant and systematic widening of the curvature around the spectral peak or break when integrating over the whole burst (see their Fig.~13). Such a widening effect must be connected to systematic variation in the BEST parameters during a burst. However, it is possible that a systematic difference between time-resolved and time-integrated parameter values for individual burst may be overwhelmed by the spread in values between different bursts, and thus not apparent in the histogram plots.

We plot in Fig.~\ref{fig:averaged} the comparisons between the averaged time-resolved BEST parameters ($\langle \alpha \rangle$, $\langle \beta \rangle$, $\langle E_\text{p} \rangle$, and $\langle E_\text{b} \rangle$) and the time-integrated BEST parameters ($\alpha^\text{int}$, $\beta^\text{int}$, $E_\text{p}^\text{int}$, and $E_\text{b}^\text{int}$) for each burst in this catalog. It can be seen that the averaged time-resolved and time-integrated $\beta$ and $E_\text{b}$ of individual bursts are consistent. The averaged time-resolved $\alpha$ are slightly harder (i.e., steeper in $\nu F_\nu$ space) than the time-integrated $\alpha$. A slight hardening of the averaged time-resolved $E_\text{p}$ is also observed. It is also observed that the standard deviations of $E_\text{p}$ are large. Moreover, it is clear from the plots that the spreads of the averaged time-resolved values are larger than their standard deviations (except for $E_\text{p}$). This implies that the time-resolved spectral behavior differs in a wide spectral range across bursts.

As \citet{Yu15b} have shown, the widening effect is primarily contributed by the high-energy side of the spectrum across the peak or break. Figure~\ref{fig:averaged} indicates that this is contributed by spectral differences that vary for individual burst, e.g., the shift in the positions of $E_\text{p}$ and the different shapes of different models (cutoff vs. broken power law). We note that, instead of $\beta$ which primarily controls the high-energy curvature in BAND and SBPL, $\alpha$ has to account for all (low-energy as well as high-energy) curvature in COMP.

\section{Summary and conclusions}
\label{sect:conc}

We present the first official gamma-ray burst time-resolved spectral catalog of the brightest subset of bursts observed by the \textit{Fermi} GBM in its first 4 years of mission. We have obtained 1,491 spectra from 81 bursts with high spectral and temporal resolution. Using a time binning criterion of S/N $=30$, it is observed that 69\% of the spectra are best fit with the Comptonized model (i.e., the high-energy cutoff power law). However, we note that this may be due to poor count statistics at the high energies, as previous catalogs have pointed out \citep[see, e.g.,][]{Kaneko06a,Goldstein12a}. Similarly, \citet{Ackermann12a} showed that for the bursts observed in GBM which happen to be in the field-of-view of the LAT but remain undetected, the upper limits are usually inconsistent with the GBM fit Band function's $\beta$, extrapolated to the LAT energy range. Whether this is a real manifestation of GRB physics or a bias due to poor high-energy count statistics, is still unclear.

We have not observed significant deviations of the distributions of fit parameters from those observed in the \textit{Fermi} GBM GRB time-integrated spectral catalogs \citep[compare Fig.~\ref{fig:BEST} to Figs.~3, 4, 6, \& 7 in][]{Gruber14a}. However, when we look at the comparison of the averaged time-resolved parameters to the time-integrated parameters, we found that the averaged time-resolved $\alpha$ and $E_\text{p}$ are harder than the time-integrated ones. Using our spectra sample, \citet{Yu15b} found that the time-integrated spectra are wider than the time-resolved spectra. This shows that while the parameter populations of all bursts as a whole show no obvious deviations between time-integrated and time-resolved results, time-integrated analysis can actually cause a widening effect, mainly due to different best-fit models used (COMP in time-resolved and BAND/SBPL in time-integrated) and the shift in the peak positions. This issue can lead to incorrect physical interpretation of, for example, the prompt emission mechanism of GRBs.

In the 4-yr GBM GRB time-integrated spectral catalog \citep{Gruber14a,vonKienlin14a}, the question of whether there is any time-resolved spectrum with very high value of $E_\text{p}$ is raised. Down to the level of the temporal resolution of the binned data sets in the current catalog, the answer is "no". The largest value of $E_\text{p}$ found in this
study is $7,409\pm597$~keV, in GRB 110721A (GBM trigger \#110721200). However, we note that very high $E_\text{p}$ on much shorter timescales cannot be excluded. \citet{Gruber14a} discussed the very high $E_\text{p} = 15 \pm 2$~MeV observed by \citet{Axelsson12a} in the "higher resolution" first time bin of GRB 110721A. Our aforementioned $E_\text{p} = 7,409\pm597$~keV is consistent at 2$\sigma$ level with their "lower resolution" first time bin of $E_\text{p} = 5,410^{+410}_{-420}$~keV.

We establish possible logical criteria for automated process of distinguishing between "hard-to-soft" and "intensity tracking" spectral evolutionary trends. With this selection scheme, only 3.5\% of bursts would be mis-attributed to the opposite kind. However, we note that inspections using human eyes are often necessary because of the existence of "hard-to-soft + intensity tracking" and FRED bursts.

We also search for plausible blackbody components in the time-resolved spectra by performing simulations on individual bursts. We find that only 3 bursts show extra blackbody components in multiple time bins. We also find that constrained fit results can be obtained only when the Planck function is added to the simple power law, using S/N $=30$ binning criterion and GBM data alone.

Finally, we note that the fact that very few blackbody emission components are found in this catalog does not necessarily imply that thermal components are in general not a dominant component for the prompt emission mechanism. There are many works recently showing that a thermal model can give rise to the observed Band shape \citep[e.g.,][]{Peer06a,Giannios08a,Peer11a,Ryde11a,Vurm11a,Lazzati13a}. Whether thermal or non-thermal emission dominates the emission mechanism of GRB prompt spectra is a hot debate topic. \citet{Yu15b} showed that all standard optically thin synchrotron emission functions are just too smooth to explain the peaks or breaks in the time-resolved spectra, and an independent conclusion is also drawn by \citet{Axelsson15a} using peak-flux spectra from the GBM time-integrated catalog. Recently, semi-empirical models \citep[e.g.,][]{Yu15a} and physical models \citep[e.g.,][]{Burgess11a,Burgess14a,Zhang15a} have been fit to time-resolved spectra of a few GRBs. In the future, direct fitting of detailed theoretical models, as oppose to empirical models, is likely the key to resolve these issues.

\begin{acknowledgements}

The authors wish to thank the anonymous referee for his/her insightful comments. HFY and JG acknowledge support by the DFG cluster of excellence "Origin and Structure of the Universe" (www.universe-cluster.de). OJR acknowledges support from Science Foundation Ireland under Grant No. 12/IP/1288. HJvE acknowledges support by the Alexander von Humboldt foundation. The GBM project is supported by the German Bundesministeriums f{\"u}r Wirtschaft und Technologie (BMWi) via the Deutsches Zentrum f{\"u}r Luft und Raumfahrt (DLR) under the contract numbers 50 QV 0301 and 50 OG 0502.

\end{acknowledgements}

\bibliographystyle{aa} 
\bibliography{mybib} 

\clearpage

\begin{appendix}

\onecolumn

\begin{landscape}

\section{Time-resolved spectral analysis results}\tiny  
\label{app:bigtable}

\setlength\LTleft{0pt}            
\setlength\LTright{0pt}           
\tabcolsep=0.11cm
\LTcapwidth=\textwidth
\begin{longtable}{cccccccccccc}
\caption{Time-resolved spectral analysis results of the BEST models. Column (1) lists the GRB names using the \textit{Fermi} GBM trigger designation. Column (2) lists the spectrum numbers within individual burst. Column (3) lists the start times $T_\text{start}$ and end times $T_\text{stop}$ for the time bins. Column (4) lists the BEST models. Columns (5) - (9) list the best-fit parameters of the BEST models, if applicable. Column (10) lists the values of CSTAT per degrees of freedom. Columns (11) and (12) list the 10~keV - 1~MeV photon and energy fluxes, respectively.}
\label{tab:bigtable} \\
\hline\hline
GRB name & spectrum & $T_\text{start}$:$T_\text{stop}$ & BEST & $A$ & $\alpha$ & $\beta$ & $E_\text{p}$ & $E_\text{b}$ & CSTAT/dof & photon flux & energy flux\\
 &  &  (s) & model & (ph~s$^{-1}$~cm$^{-2}$~keV$^{-1}$) &  &  & (keV) & (keV) &  & (ph~s$^{-1}$~cm$^{-2}$) & (erg~s$^{-1}$~cm$^{-2}$)\\
(1) & (2) & (3) & (4) & (5) & (6) & (7) & (8) & (9) & (10) & (11) & (12)\\
\hline
\endfirsthead
\multicolumn{10}{c}%
{{\bfseries \tablename\ \thetable{} -- continued from previous page}} \\
\hline\hline
GRB name & spectrum & $T_\text{start}$:$T_\text{stop}$ & BEST & $A$ & $\alpha$ & $\beta$ & $E_\text{p}$ & $E_\text{b}$ & CSTAT/dof & photon flux & energy flux\\
 &  &  (s) & model & (ph~s$^{-1}$~cm$^{-2}$~keV$^{-1}$) &  &  & (keV) & (keV) &  & (ph~s$^{-1}$~cm$^{-2}$) & (erg~s$^{-1}$~cm$^{-2}$)\\
(1) & (2) & (3) & (4) & (5) & (6) & (7) & (8) & (9) & (10) & (11) & (12)\\
\hline
\endhead
\hline \multicolumn{10}{c}{\textit{Continued on next page}} \\
\endfoot
\hline
\endlastfoot
080817161 & 1 & -6.144:4.344 & COMP & $0.0067\pm0.0010$ & $-0.835\pm0.060$ & - & $637.80\pm96.70$ & - & 570.4/480 & 2.1682$\pm$0.24 & 5.7087E-07$\pm$5.5E-08\\
080817161 & 2 & 4.344:5.885 & COMP & $0.0313\pm0.0034$ & $-0.829\pm0.047$ & - & $668.50\pm79.10$ & - & 492.51/480 & 10.172$\pm$0.82 & 2.7450E-06$\pm$1.8E-07\\
080817161 & 3 & 5.885:7.150 & COMP & $0.0444\pm0.0053$ & $-0.827\pm0.048$ & - & $512.90\pm53.90$ & - & 529.13/480 & 13.263$\pm$1.1 & 3.1743E-06$\pm$2.2E-07\\
080817161 & 4 & 7.150:8.200 & SBPL & $0.0439\pm0.0048$ & $-0.863\pm0.056$ & $-2.121\pm0.122$ & $388.69\pm137.52$ & $191.10\pm30.90$ & 517.04/479 & 15.266$\pm$1.4 & 3.3723E-06$\pm$2.6E-07\\
080817161 & 5 & 8.200:10.956 & COMP & $0.0303\pm0.0049$ & $-0.871\pm0.057$ & - & $348.20\pm36.10$ & - & 522.13/480 & 8.1164$\pm$0.93 & 1.5151E-06$\pm$1.4E-07\\
080817161 & 6 & 10.956:12.591 & COMP & $0.0288\pm0.0041$ & $-0.865\pm0.052$ & - & $479.60\pm56.60$ & - & 497.74/480 & 8.5970$\pm$0.92 & 1.9294E-06$\pm$1.7E-07\\
080817161 & 7 & 12.591:13.934 & COMP & $0.0380\pm0.0050$ & $-0.859\pm0.051$ & - & $482.40\pm54.60$ & - & 491.33/480 & 11.319$\pm$1.1 & 2.5594E-06$\pm$2.0E-07\\
080817161 & 8 & 13.934:15.019 & COMP & $0.0407\pm0.0048$ & $-0.865\pm0.047$ & - & $591.50\pm69.80$ & - & 547.03/480 & 12.969$\pm$1.1 & 3.2167E-06$\pm$2.3E-07\\
\hline
\end{longtable}

\clearpage

\section{$E_\text{p}$ evolutionary trends}

\begin{longtable}{ccccccccccccccc}
\caption{$E_\text{p}$ evolutionary trends. Column (1) lists the GRB names. Column (2) lists the numbers of spectra with $E_\text{p}$. Columns (3), (7), and (11) list the Spearman's Rank Correlation Coefficients between $E_\text{p}$ and the photon flux and the 90\%, 95\%, and 99\% confidence intervals, respectively. Columns (4), (8), and (12) list the Spearman's Rank Correlation Coefficients between $E_\text{p}$ and the energy flux and the 90\%, 95\%, and 99\% confidence intervals, respectively. Columns (5), (9), and (13) list the Spearman's Rank Correlation Coefficients between $E_\text{p}$ and the time and the 90\%, 95\%, and 99\% confidence intervals, respectively. Columns (6), (10), and (14) list the trends as determined by the computer for the 90\%, 95\%, and 99\% confidence intervals, respectively. Column (15) lists the trends as determined by human eyes.}
\label{tab:trends} \\
\hline\hline
GRB name & $N$ & $\rho_\text{ph}(90\%)$ & $\rho_\text{en}(90\%)$ & $\rho_\text{t}(90\%)$ & trend(90\%) & $\rho_\text{ph}(95\%)$ & $\rho_\text{en}(95\%)$ & $\rho_\text{t}(95\%)$ & trend(95\%) & $\rho_\text{ph}(99\%)$ & $\rho_\text{en}(99\%)$ & $\rho_\text{t}(99\%)$ & trend(99\%) & trend(by eyes)\\
(1) & (2) & (3) & (4) & (5) & (6) & (7) & (8) & (9) & (10) & (11) & (12) & (13) & (14) & (15)\\
\hline
\endfirsthead
\multicolumn{15}{c}%
{{\bfseries \tablename\ \thetable{} -- continued from previous page}} \\
\hline\hline
GRB name & $N$ & $\rho_\text{ph}(90\%)$ & $\rho_\text{en}(90\%)$ & $\rho_\text{t}(90\%)$ & trend(90\%) & $\rho_\text{ph}(95\%)$ & $\rho_\text{en}(95\%)$ & $\rho_\text{t}(95\%)$ & trend(95\%) & $\rho_\text{ph}(99\%)$ & $\rho_\text{en}(99\%)$ & $\rho_\text{t}(99\%)$ & trend(99\%) & trend(by eyes)\\
(1) & (2) & (3) & (4) & (5) & (6) & (7) & (8) & (9) & (10) & (11) & (12) & (13) & (14) & (15)\\
\hline
\endhead
\hline \multicolumn{15}{c}{\textit{Continued on next page}} \\
\endfoot
\hline
\endlastfoot
080817161 & 15 & $0.354^{+0.335}_{-0.459}$ & $0.546^{+0.250}_{-0.409}$ & $-0.700^{+0.327}_{-0.172}$ & undeter. & $0.354^{+0.379}_{-0.547}$ & $0.546^{+0.281}_{-0.499}$ & $-0.700^{+0.407}_{-0.192}$ & undeter. & $0.354^{+0.451}_{-0.711}$ & $0.546^{+0.329}_{-0.676}$ & $-0.700^{+0.577}_{-0.223}$ & h.t.s. & h.t.s.+in.track.\\
080825593 & 12 & $0.322^{+0.386}_{-0.533}$ & $0.538^{+0.279}_{-0.485}$ & $-0.832^{+0.263}_{-0.108}$ & undeter. & $0.322^{+0.434}_{-0.631}$ & $0.538^{+0.311}_{-0.590}$ & $-0.832^{+0.338}_{-0.119}$ & h.t.s. & $0.322^{+0.509}_{-0.803}$ & $0.538^{+0.359}_{-0.789}$ & $-0.832^{+0.508}_{-0.135}$ & h.t.s. & h.t.s.\\
080916009 & 21 & $0.168^{+0.338}_{-0.383}$ & $0.249^{+0.317}_{-0.382}$ & $-0.288^{+0.379}_{-0.306}$ & undeter. & $0.168^{+0.391}_{-0.452}$ & $0.249^{+0.366}_{-0.454}$ & $-0.288^{+0.452}_{-0.352}$ & undeter. & $0.168^{+0.483}_{-0.579}$ & $0.249^{+0.448}_{-0.588}$ & $-0.288^{+0.589}_{-0.430}$ & undeter. & undeter.\\
081009140 & 0 & - & - & - & - & - & - & - & - & - & - & - & - & -\\
081124060 & 0 & - & - & - & - & - & - & - & - & - & - & - & - & -\\
081125496 & 4 & - & - & - & - & - & - & - & - & - & - & - & - & -\\
081207680 & 9 & $0.683^{+0.223}_{-0.521}$ & $0.717^{+0.201}_{-0.491}$ & $0.367^{+0.418}_{-0.646}$ & in.track. & $0.683^{+0.244}_{-0.648}$ & $0.717^{+0.219}_{-0.616}$ & $0.367^{+0.462}_{-0.760}$ & in.track. & $0.683^{+0.272}_{-0.896}$ & $0.717^{+0.244}_{-0.866}$ & $0.367^{+0.526}_{-0.949}$ & undeter. & in.track.\\
081215784 & 16 & $0.453^{+0.284}_{-0.421}$ & $0.538^{+0.247}_{-0.394}$ & $-0.650^{+0.341}_{-0.193}$ & undeter. & $0.453^{+0.322}_{-0.508}$ & $0.538^{+0.278}_{-0.480}$ & $-0.650^{+0.422}_{-0.217}$ & undeter. & $0.453^{+0.381}_{-0.675}$ & $0.538^{+0.328}_{-0.650}$ & $-0.650^{+0.589}_{-0.253}$ & h.t.s. & h.t.s.+in.track.\\
\hline
\end{longtable}

\end{landscape}

\clearpage

\section{Identified significant blackbodies}

\setlength\LTleft{0pt}            
\setlength\LTright{0pt}           
\tabcolsep=0.11cm
\LTcapwidth=\textwidth
\begin{longtable}{cccccccc}
\caption{Best-fit parameters of the identified significant blackbodies of the PLBB model. Column (1) lists the GRB names. Column (2) lists the spectrum numbers within individual burst. Column (3) lists the start times $T_\text{start}$ and end times $T_\text{stop}$ for the time bins. Columns(4) - (7) list the best-fit parameters of the PLBB model. Column (7) lists the values of $\Delta\text{CSTAT} = \text{CSTAT(BEST)} - \text{CSTAT(PLBB)}$. $^\text{a}$We note that for these 2 spectra the BEST models are SBPL (the rest BEST models from this burst are all COMP), CSTAT values between SBPL and PLBB are directly compared without performing any simulations.}
\label{tab:blackbodies} \\
\hline\hline
GRB name & spectrum & $T_\text{start}$:$T_\text{stop}$ & $A_\text{PL}$ & $\alpha$ & $A_\text{BB}$ & $kT$ & $\Delta\text{CSTAT}$ \\
 &  & (s) & (ph~s$^{-1}$~cm$^{-2}$~keV$^{-1}$) &  & (ph~s$^{-1}$~cm$^{-2}$~keV$^{-1}$) & (keV) &  \\
(1) & (2) & (3) & (4) & (5) & (6) & (7) & (8) \\
\hline
\endfirsthead
\multicolumn{8}{c}%
{{\bfseries \tablename\ \thetable{} -- continued from previous page}} \\
\hline\hline
GRB name & spectrum & $T_\text{start}$:$T_\text{stop}$ & $A_\text{PL}$ & $\alpha$ & $A_\text{BB}$ & $kT$ & $\Delta\text{CSTAT}$ \\
 &  & (s) & (ph~s$^{-1}$~cm$^{-2}$~keV$^{-1}$) &  & (ph~s$^{-1}$~cm$^{-2}$~keV$^{-1}$) & (keV) &  \\
(1) & (2) & (3) & (4) & (5) & (6) & (7) & (8) \\
\hline
\endhead
\hline \multicolumn{8}{c}{\textit{Continued on next page}} \\
\endfoot
\hline
\endlastfoot
090618353 & 46 & 79.706:80.155 & $0.0249\pm0.0061$ & $-1.70\pm0.06$ & $3.135E-04\pm1.22E-04$ & $22.12\pm1.72$ & 57.57\\
090618353 & 77 & 96.641:110.742 & $0.0051\pm0.0010$ & $-1.88\pm0.03$ & $1.333E-04\pm3.59E-05$ & $15.59\pm0.81$ & 128.1\\
090902462$^\text{a}$ & 10 & 6.433:6.851 & $0.0216\pm0.0021$ & $-1.56\pm0.05$ & $1.456E-06\pm1.79E-07$ & $181.20\pm6.14$ & 12.43\\
090902462$^\text{a}$ & 11 & 6.851:7.218 & $0.0233\pm0.0024$ & $-1.58\pm0.06$ & $1.600E-06\pm2.08E-07$ & $175.70\pm6.18$ & 29.47\\
090902462 & 12 & 7.218:7.500 & $0.0315\pm0.0031$ & $-1.61\pm0.05$ & $2.050E-06\pm2.78E-07$ & $170.50\pm6.25$ & 82.4\\
090902462 & 13 & 7.500:7.774 & $0.0387\pm0.0037$ & $-1.62\pm0.05$ & $9.601E-07\pm1.28E-07$ & $222.10\pm8.40$ & 59.65\\
090902462 & 14 & 7.774:7.958 & $0.0516\pm0.0048$ & $-1.63\pm0.05$ & $9.149E-07\pm1.48E-07$ & $227.70\pm10.50$ & 102.04\\
090902462 & 15 & 7.958:8.119 & $0.0804\pm0.0069$ & $-1.65\pm0.04$ & $9.125E-07\pm1.60E-07$ & $228.50\pm11.60$ & 67.74\\
\hline
\end{longtable}

\twocolumn

\section{Connection to time-integrated catalogs: the GOOD sample}
\label{App:goodsample}

\begin{figure*}
\resizebox{\hsize}{!}
{\includegraphics[width = 18 cm]{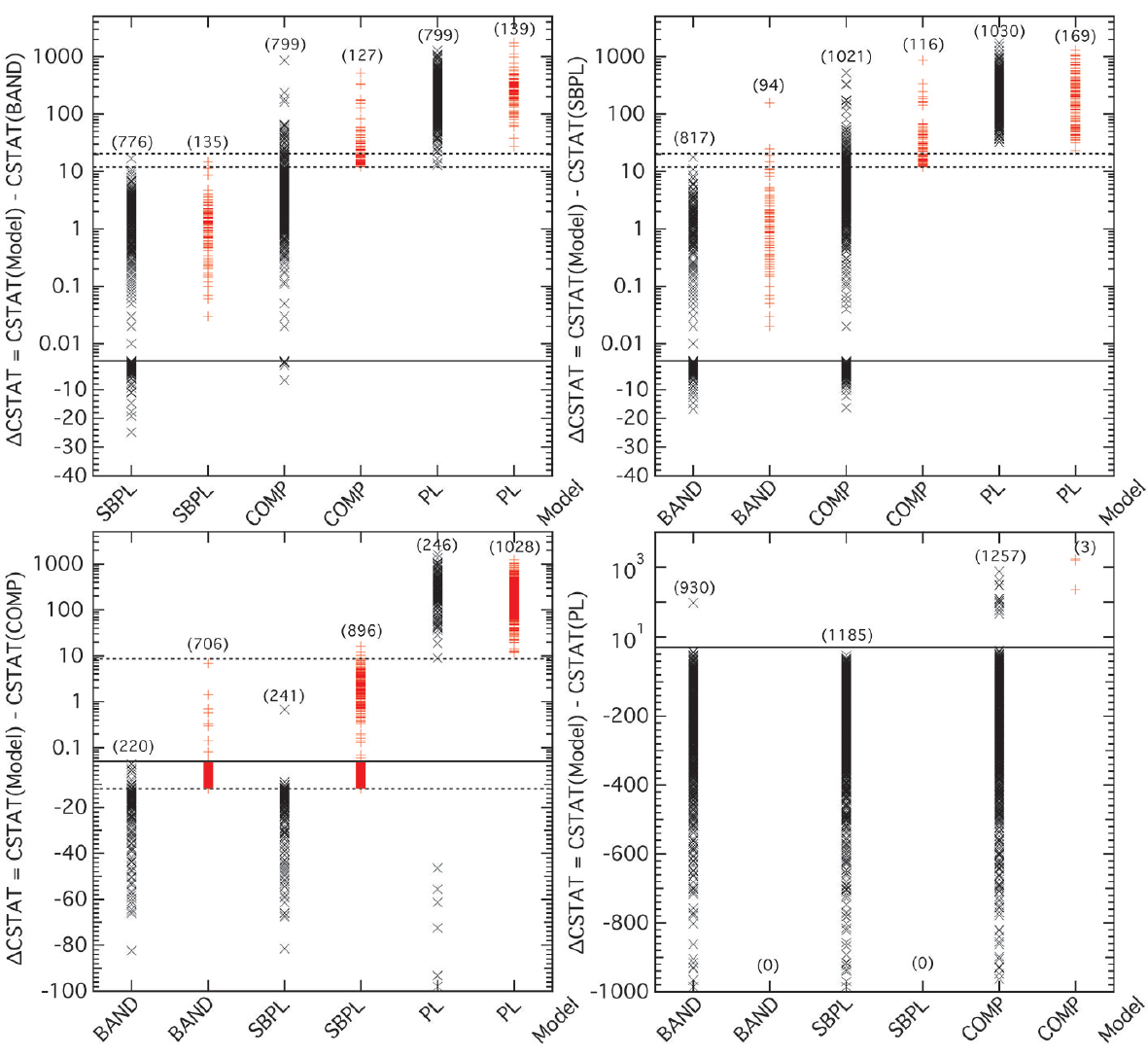}}
\caption{Difference between the CSTAT values for every pairs of GOOD fits from each spectrum. The top left, top right, bottom left, and bottom right panel shows the CSTAT differences, $\Delta$CSTAT, between the other models and BAND, SBPL, COMP, and PL, respectively. Red crosses indicate the latter model is the BEST model (i.e., also a GOOD model), and black crosses indicate the latter model is the GOOD-but-not-BEST model. The numbers in brackets indicate the numbers of pairs of models compared (those spectra with only one GOOD and/or BEST fits were not compared). The two dashed lines indicate 11.83 and $11.83+8.58=20.41$ in the top panels, and $-11.83$ and 8.58 in the bottom left panel. The solid line separates the logarithmic positive $y$-axis and linear negative $y$-axis. One black cross for CSTAT(PL) $-$ CSTAT(BAND) = $-95$, about 20 black crosses for CSTAT(SBPL) $-$ CSTAT(COMP) and CSTAT(PL) $-$ CSTAT(COMP) down to $\sim -1700$, and about 20 black crosses for CSTAT(Model) $-$ CSTAT(PL) down to $\sim -6800$ are not shown for clear display purpose.}
\label{fig:check}
\end{figure*}

In the first two GBM GRB time-integrated spectral catalogs \citep{Goldstein12a,Gruber14a}, a GOOD sample was defined. We do not show the GOOD sample statistics in this catalog, because we found that the GOOD criteria do not guarantee good fits. We investigate this effect here in the current appendix section.

Since the definition of the GOOD sample does not include any goodness-of-fit measure, it is necessary to investigate the performance of the fits w.r.t. the data, manifested by the CSTAT values. In Fig.~\ref{fig:check} we plot the differences between the CSTAT values, $\Delta$CSTAT, for every pair of GOOD fits for each spectrum. The top left, top right, bottom left, and bottom right panel show the $\Delta$CSTAT between the other models and BAND, SBPL, COMP, and PL, respectively. For BAND and SBPL, it can be seen that when they are GOOD but not BEST, their $\Delta$CSTATs are $\sim 0.5$ - 50 comparing to COMP (a 3-parameters model) and the other 4-parameters models, and even larger ($\sim 100$) comparing to PL (a 2-parameters model). This indicates that the GOOD BAND and SBPL are generally reliable good fits. The $\Delta$CSTAT of the GOOD-but-not-BEST COMP are concentrated $\sim-60$ to $-10$, and that for the BEST fits are within $-11.83$ to 10, indicating that the GOOD-but-not-BEST COMP do not perform as well as the BEST COMP w.r.t. data.

However, looking at the $\Delta$CSTAT of PL, we can immediately see that most of the GOOD-but-not-BEST PL have very negative values. This indicates that the GOOD statistics of PL's $\alpha$ are not reliable. Moreover, in almost all of the BEST PL cases, PL becomes BEST by default, as there are no other models that lead to GOOD fits. While this issue does not necessarily imply bad description of the data by PL, we suggest researchers always perform careful inspection of the PL fit results. Nevertheless, this issue does not affect the current catalog results, because all analyses are done using the BEST sample and do not include PL fits.

\end{appendix}

\end{document}